\documentclass[12pt]{article}
\usepackage{a4}
\usepackage{psfig}
\usepackage{amssymb}

\newcommand{\half}{\frac{1}{2}}
\newcommand{\be}{\begin{equation}}
\newcommand{\ee}{\end{equation}}
\newcommand{\bean}{\begin{eqnarray*}}
\newcommand{\eean}{\end{eqnarray*}}
\newcommand{\bea}{\begin{eqnarray}}
\newcommand{\eea}{\end{eqnarray}}

\newcommand{\bra}{\langle}
\newcommand{\ket}{\rangle}
\newcommand{\Tr}{{\rm Tr}}

\newcommand{\dslash}{/\!\!\! \partial}
\newcommand{\Dslash}{/\!\!\!\! D}
\newcommand{\calDslash}{/\!\!\!\! {\cal D}}
\newcommand{\Aslash}{/\!\!\!\! A}

\newcommand{\CC}{{\cal C}}
\newcommand{\HD}{{{\cal H}_D}}

\newcommand{\al}{\alpha}
\newcommand{\bt}{\beta}
\newcommand{\gm}{\gamma}
\newcommand{\dl}{\delta}
\newcommand{\ep}{\epsilon}

\newcommand{\lm}{\lambda}
\newcommand{\rh}{\rho}
\newcommand{\sg}{\sigma}
\newcommand{\ta}{\tau}
\newcommand{\ph}{\phi}

\newcommand{\ch}{\chi}
\newcommand{\ps}{\psi}
\newcommand{\om}{\omega}

\newcommand{\Gm}{\Gamma}

\newcommand{\Ph}{\Phi}

\newcommand{\NPB}[1]{{\em Nucl. Phys.} {\bf B#1}}
\newcommand{\PLB}[1]{{\em Phys. Lett.} {\bf B#1}}
\newcommand{\PRD}[1]{{\em Phys. Rev.} {\bf D#1}}
\newcommand{\PRL}[1]{{\em Phys. Rev. Lett.} {\bf #1}}
\newcommand{\MPLA}[1]{{\em Mod. Phys. Lett.} {\bf A#1}}

\begin{document}

\title{
\vskip -100pt
{
\begin{normalsize}
\mbox{} \hfill THU-98-45\\
\mbox{} \hfill ITFA-98-31\\
\mbox{} \hfill hep-ph/9812413\\
\mbox{} \hfill December 1998\\
\vskip  70pt
\end{normalsize}
}
Real-Time Dynamics with Fermions on a Lattice
}

\author{   
Gert Aarts$^a$\thanks{email: aarts@phys.uu.nl}
\addtocounter{footnote}{1}
and
Jan Smit$^{a,b}$\thanks{email: jsmit@phys.uva.nl}\\
\normalsize
{\em $\mbox{}^a$Institute for Theoretical Physics, Utrecht University}\\
\normalsize
{\em            Princetonplein 5, 3584 CC Utrecht, the Netherlands}\\
\normalsize
{\em $\mbox{}^b$Institute for Theoretical Physics, University of 
Amsterdam}\\
\normalsize
{\em            Valckenierstraat 65, 1018 XE Amsterdam, the Netherlands}
\normalsize
}
\maketitle
 
\renewcommand{\abstractname}{\normalsize Abstract}
\begin{abstract}
\normalsize
The $1+1$ dimensional abelian Higgs model with fermions is
a toy model for the theory of electroweak baryogenesis.
We study the dynamics of the model with axially coupled fermions
in real-time. The model is defined  on a spacetime lattice to preserve 
gauge invariance and to obtain numerical stability in a simple way. 
We take into account the phenomenon of lattice fermion doubling.
The dynamics is approximated by treating the inhomogeneous Bose fields
classically, which is justified in a large $N_f$ approximation. 
The back reaction on the Bose fields due to fermion field fluctuations is 
calculated using a mode function expansion.
We discuss and present numerical results for the response of fermions to 
sphaleron transitions, the renormalizability of the effective equations 
of motion and non-perturbative dynamics in the framework of nonequilibrium 
quantum field theory. 
The long-time behaviour of the system is discussed and we speculate about 
applications to finite density calculations.

\end{abstract}

\newpage

\renewcommand{\theequation}{\arabic{section}.\arabic{equation}}

\section{Introduction}
\setcounter{equation}{0}

The time evolution of quantum fields in and out of equilibrium plays an 
important role in the early universe and in heavy ion 
collisions.
Detailed questions concerning baryogenesis, inflation, the chiral phase 
transition and time evolution during phase transitions in 
general, require a non-perturbative analysis, preferably directly in real 
(as opposed to imaginary) time. 
Since exact answers are usually not available, a 
considerable amount of work has been devoted to finding suitable
approximation methods.

One approach which has received a lot of attention recently is 
based on Hartree or large $N$ approximations \cite{CoMo87, CoMo94, Boya96}, 
and has been applied to
pair production in strong electric fields \cite{pairpro}, (inflationary) 
cosmology \cite{cosmo, baacke, fermionnumerics}, the formation of chiral 
condensates in 
heavy ion collisions \cite{chcon} and nonequilibrium symmetry breaking 
\cite{nonequi}.
In this approximation method, the fields are written as the sum of 
`classical' or `background' 
fields (which are then called order parameter, or 
condensate\footnote{We shall use the terms classical field, background 
field and order parameter interchangeably.}) and quantum 
fluctuations around them. A set of coupled equations for 
the background fields and the fluctuations can be derived from 
first principles in several ways from the Schwinger-Keldysh formalism for 
real-time (non-)equilibrium field theory, and they have the form of 
semiclassical equations of motion. A nice property 
of the equations is that they contain only the usual quantum divergences  
which can be renormalized with the usual counterterms. 

However, in numerical work the emphasis in the literature has been mainly on 
situations where the classical field is homogeneous. 
Although this is sufficient for some situations 
(inflationary scenarios, pair creation in strong homogeneous electric 
fields), this is certainly not the case for 
sphaleron transitions or realistic treatments of heavy ion collisions.

A useful toy model 
for obtaining a better understanding of the dynamics that plays a role 
in various scenarios for baryogenesis
 (for recent reviews, see, e.g., \cite{baryo}) 
has been the abelian Higgs model in $1+1$ dimensions 
 \cite{abh,TaSm98}. 
The model has many properties in common with the electroweak sector of the 
Standard Model, in particular it has a non-trivial vacuum 
structure which is labeled with the Chern-Simons number, 
and sphaleron barriers separating the vacua. 
Furthermore, extending it
with axially coupled fermions results 
into a model with anomalous fermion
number non-conservation, similar to baryon number violation in the 
Standard Model.

In this paper we study this abelian Higgs model with axial fermions, 
using the nonequilibrium method mentioned above. We treat the bosonic 
fields as classical background fields, and the 
fermions as fluctuations, which is justified for $N_f$ fermions 
in the limit of large $N_f$. 
The bosonic fields are not restricted to be homogeneous, allowing for 
genuine sphaleron transitions. The equations of motion for the 
bosonic fields have the form of classical field equations coupled to the 
fermion current or force. The fermions are treated with a mode 
function expansion, in which the mode functions obey a Dirac-like 
equation in the presence of inhomogeneous background fields. 

We formulate the model on a lattice in space and time.
This allows for a simple and stable time 
integration of the coupled partial differential equations for finite 
lattice spacing. Furthermore, gauge invariance can be implemented
and exact symmetries of the lattice action
give rise to exact conservation laws at finite lattice spacing.

In the literature, to our knowledge fermion fields have been 
treated numerically only in the case of homogeneous mean fields 
\cite{pairpro,fermionnumerics} (pair production, preheating in inflation), 
while fermions in the presence of inhomogeneous mean fields have sofar been 
investigated analytically only (see e.g. \cite{fermionanalytic}).

In the remainder of this Introduction, we give the outline of the
paper. In section \ref{seccontinuum}, we introduce and 
 summarize some basic 
properties of the abelian Higgs model with axial fermions in the 
continuum. In particular, we recall that the axial gauge symmetry can be 
transformed into a vector one, which is convenient because it allows us 
to extend known methods for fermions on the lattice in 
imaginary-time to real-time. This transformation leads to an unusual 
Majorana-type Yukawa coupling, hence, we go from a formulation with Dirac 
spinors to one with Majorana spinors.
 The next two sections are rather technical. 
In section \ref{seclattice}  we present the model on 
the lattice. Fermion fields on a lattice lead to the 
fermion doubling phenomenon \cite{Wil75,KarSm81,books},
and we describe in this section how we deal with this.
Symmetries, currents and useful observables are discussed in section
\ref{secsymmetries}. The large $N_f$ effective action is briefly derived 
in section \ref{secaction} and the resulting equations of motion are 
given in section \ref{secequations}. 
The effective equations of motion are solved numerically and we 
present results in sections \ref{sechandmade}-\ref{secthermalbose}.
In section \ref{sechandmade} we use handmade time dependent 
Bose field configurations that go 
from one vacuum to another via a sphaleron transition. These external
fields are used to test the properties of the fermion system.
Issues related to renormalization 
are discussed in section \ref{secrenorm}.
In section \ref{secnumerics}, we present numerical results for 
the full non-perturbative dynamics, including the fermion back reaction on 
the Bose fields. 
We discuss the long-time behaviour of the 
system in section \ref{secthermalbose} and also mention a possible 
application to finite density. We end with a summary
and outlook.

Conventions and technicalities are summarized  in several appendices. 
Appendix \ref{appconventions} contains our conventions and
in appendix \ref{appinitialf} we give the initial mode functions used for the
fermions.
Finally, in appendix \ref{appspec} we show how the relation between 
spectral flow and the anomalous conservation law works out in our lattice 
model.

\section{Continuum model}
\label{seccontinuum}
\setcounter{equation}{0}

We consider the abelian Higgs model in $1+1$ dimensions with 
axially coupled fermions. The action consists of a gauge, Higgs and fermion 
part and is given, in the continuum, by
\be
\label{eqaction} 
S = S_G + S_H + S_F,\ee
with
\bea
\label{eqactiong}
 S_G &=& -\int d^2 x \frac{1}{4e^2}F_{\mu\nu}F^{\mu\nu},\\
\label{eqactionh}
 S_H &=& -\int d^2 x \left[ (D_\mu \phi)^*D^\mu \phi + V(\phi)\right],
\;\;\;\;
 V(\phi) =   \lambda\left(\phi^*\phi - v^2/2\right)^2,\\
S_F &=& -\int d^2x \left[ \bar\psi(\dslash +  iq\Aslash\gm_5)\psi + G
\bar\psi(\phi^*P_L+\phi P_R)\psi\right]. 
\eea
Here $F_{\mu\nu} = \partial_\mu A_\nu-\partial_\nu A_\mu$, $D_\mu = 
\partial_\mu - iA_\mu$, and $P_{R,L} = (1\pm\gm_5)/2$.
The model is defined on a circle with circumference $L$ in the spatial 
direction. The bosonic (fermionic) fields 
obey (anti)periodic boundary conditions.
Our conventions are summarized in appendix \ref{appconventions}.

The model is invariant under local axial gauge transformations
\[
 \psi(x) \to e^{iq\xi(x)\gm_5}\psi(x),\;\;\;\;
 \phi(x) \to e^{-i\xi(x)}\phi(x), \;\;\;\; A_\mu(x) \to A_\mu(x) 
-\partial_\mu \xi(x).
\] 
Gauge invariance implies that the fermions have one half of the charge of 
the scalar field, $q=\half$. The conserved (axial) current, $\partial_\mu 
j^\mu=0$, associated to the global gauge symmetry, is 
\[ j^\mu = -j^\mu_5 + j^\mu_h, \;\;\;\;
j^\mu_5 = iq\bar\psi\gm^\mu\gm_5\psi,\;\;\;\;
j^\mu_h =  i(D^\mu\phi)^*\phi -i\phi^*D^\mu\phi.
\]
Furthermore there is a global vector symmetry with a classically conserved 
current 
\[ \psi \to e^{i\om}\psi, \;\;\;\; j^\mu_f = i\bar\psi\gm^\mu\psi,\] 
corresponding to fermion number.

 The global vector symmetry is anomalous,
and the anomaly equation reads
\[ \partial_\mu j^\mu_f = -q_{\rm top}.\]
The topological charge density can be written as  the divergence of the 
Chern-Simons current
\[
q_{\rm top} = \frac{1}{4\pi}\ep^{\mu\nu}F_{\mu\nu} = \partial_\mu 
C^\mu,\;\;\;\;
C^\mu = \frac{1}{2\pi}\ep^{\mu\nu}A_\nu.\]
Fermion number violation is expressed as
 \be
Q_f(t)-Q_f(0) = -(C(t)-C(0)),
\ee
relating the total fermion number and the Chern-Simons number 
\be
Q_f(t)=\int dx^1 j^0_f(x),\;\;\;\;C(t)=\int dx^1 C^0(x)=-\frac{1}{2\pi}\int 
dx^1 A_1(x). 
\ee

It is possible to transform the model to one with a vector gauge 
symmetry, by performing charge conjugation only on the right-handed fields
 \cite{Bock}. 
This is convenient, because we want to treat the model on a 
lattice.\footnote{A similar transformation 
can also be performed in the electroweak SU(2) Higgs model in $3+1$ 
dimensions.}
The transformation is explicitly
\be
\label{eqtrafo}
\psi_R = (\bar\psi_R' {\cal C})^T, 
\;\;\;\;
\bar\psi_R = -({\cal C}^\dagger\psi_R')^T,
\;\;\;\;
\psi_L = \psi_L',\;\;\;\;\bar\psi_L = \bar \psi_L'.
\ee
${\cal C}$ denotes the charge conjugation matrix.
In these primed variables, the fermion part of the action is
\be
\label{eqactionf}
S_F = -\int d^2x\left[ 
\bar\psi' (\dslash -iq\Aslash)\psi' + 
\half G\psi'^T{\cal C}^\dagger\phi^*\psi' - 
\half G\bar\psi'{\cal C}\phi\bar\psi'^T\right].
\ee
In the Yukawa mass term only the antisymmetric part survives. 
The gauge-fermion part of the action, with $G=0$, has now the form of the 
Schwinger model, or QED in $1+1$ dimensions, with $q=\half$ charged 
fermions. The Yukawa term, coupling 
the fermion field to the scalar field, has turned into a Majorana-like mass 
term. 

In terms of the primed fermion field, the gauge symmetry is 
vector-like 
\[ \psi'(x)\to e^{-iq\xi(x)}\psi'(x),\;\;\;\; \phi(x) \to 
e^{-i\xi(x)}\phi(x),
\;\;\;\; A_\mu(x) \to A_\mu(x) -\partial_\mu \xi(x),
\]
 with the conserved current
\[ j^\mu = j'^\mu_f + j^\mu_h, \;\;\;\;j'^\mu_f = 
iq\bar\psi'\gm^\mu\psi' = -j^\mu_5.\] 
The global symmetry that involves only the fermion field is now axial, 
with a classically conserved current 
\be
\label{eqgm5cont}
\psi' \to e^{-i\om\gm_5}\psi', \;\;\;\; 
j'^\mu_5 = i\bar\psi'\gm^\mu\gm_5\psi' = -j^\mu_f.\ee
The anomaly equation reads
 \be
Q'_5(t)-Q'_5(0) = C(t)-C(0),\;\;\;\;Q'_5(t)=\int dx^1 j'^0_5(x).
\ee
In the rest of this paper we will only work with the transformed action. 
Therefore we drop the primes from the fermion fields
and charges in the following sections.

In the remainder of this section, we summarize some 
 properties of the model. 
We work in the temporal gauge $A_0=0$ throughout the paper. 
The bosonic vacuum can
be labeled with the Chern-Simons number $C$. The local energy minima are 
characterized by $C$ being integer valued and $|\phi|=v/\sqrt{2}$. The 
vacua are separated by finite energy sphaleron barriers, where $C$ is 
half-integer. 
The height of the energy barrier is given by the sphaleron energy $E_{\rm 
sph} = \frac{2}{3}\sqrt{2\lambda} v^3$ \cite{BochShap87}. 
We will show the vacuum structure in more detail, also in the presence of 
fermions, 
when we use handmade sphaleron transitions in section \ref{sechandmade}.

The dimensions of the fields and coupling constants is as follows:
$\phi$ and $v$ are dimensionless, $\psi$ has a mass dimension of 1/2, 
$A_\mu, e$ and $G$ have dimension 1, and $\lambda$ has dimension 2.

Elementary powercounting shows that besides the partition function, 
only the one loop self energy of the scalar field is superficially 
divergent. This divergence can be canceled with the appropiate choice 
for the bare scalar mass (or $v^2$).

Finally,
the tree level masses, due to classical Higgs expectation value, 
 of respectively the scalar, gauge and fermion fields, are
\[  m_H^2 = 2\lambda v^2,\;\;\;\; m_{A, h}^2 = e^2v^2,\;\;\;\;
m_F^2=\frac{G^2v^2}{2},\;\;\;\;\mbox{(tree level)}.\]
At the one loop level, the fermion field not only affects the scalar self 
energy, it contributes also to the mass of the 
gauge boson
\[  m_{A, f}^2 = \frac{q^2e^2}{\pi}, \;\;\;\;\mbox{(one loop fermion)}.\] 

This concludes the introduction of the model. 
In the next section, we discretize it on a lattice in space and time.

\section{Lattice formulation}
\label{seclattice}
\setcounter{equation}{0}

In section \ref{secaction} and beyond, we derive and study a set of 
effective equations of motion, that involve the Bose fields and the 
back reaction of the fermion field. This study is done numerically, and 
involves a discretization. A convenient way to discretize the equations 
is by use of a lattice formulation in space and real time. There are good 
reasons to use a lattice: exact symmetries of the lattice action give 
rise to exactly conserved quantities (i.e. for finite lattice 
spacing), the lattice distance is a gauge invariant ultraviolet cutoff, 
the resulting time integration algorithm is simple and stable.
 For a general introduction to lattice gauge theories in euclidean 
spacetime, we refer to \cite{books}.
Before we write down the effective equations, we first discuss in this 
section the lattice formulation and in the next section exact symmetries 
and conserved currents.

The model is defined on a lattice in space and (real) time, $x=(x^0, 
x^1)$. When there is no confusion we also use $x^0=t, x^1=x$.
The lattice spacing in direction $\mu$ is denoted by
$a_\mu = (a_0, a_1\equiv a)$, and there are $N$ spatial 
lattice points ($L=aN$). As indicated in the previous section, from 
now on we only work with the transformed action with a vector-like gauge 
symmetry (i.e. (\ref{eqaction}-\ref{eqactionh}, \ref{eqactionf}) in  
the continuum), hence we drop the primes from the 
fermion fields and currents. 

To put the Bose fields on the lattice is straightforward. We replace 
the (covariant) derivative by the lattice (covariant) derivative as
\bean
\partial_\mu f(x) = \frac{1}{a_\mu}\left[f(x+\hat a_\mu) - 
f(x)\right],
&&\;\; 
D_\mu f(x) = \frac{1}{a_\mu}\left[U_{\mu}(x)f(x+\hat a_\mu)-f(x)\right],\\
\partial'_\mu f(x) = \frac{1}{a_\mu}\left[f(x) - f(x-\hat 
a_\mu)\right],
&&\;\;
D'_\mu f(x) = \frac{1}{a_\mu}\left[f(x) - 
U^*_{\mu}(x-\hat a_\mu)f(x-\hat a_\mu)\right].
\eean
$\hat a_\mu$ denotes the unit vector in the $\mu$ direction, and we 
use the notation
\[ U_{\mu}(x) = e^{-ia_\mu A_{\mu}(x)}.\]
The gauge and scalar part of the action keep the same form as in the 
continuum 
\bea
\label{eqactionglat}
 S_G &=& -\sum_{x,\mu,\nu} \frac{1}{4e^2}F_{\mu\nu}F^{\mu\nu},\\
\label{eqactionhlat}
 S_H &=& -\sum_{x,\mu} (D_\mu \phi)^*D^\mu \phi - \sum_x V(\phi).
\eea
We use the non-compact formulation for the gauge field.

To put fermions on a lattice is less straightforward. The naive way leads 
to so-called fermion doublers in space and time 
 \cite{Wil75,KarSm81}.
Instead of one physical fermion 
particle, the lattice action describes, close 
enough to the continuum limit, $2^2$ physical fermions.
One effect 
of the doublers is to cancel the anomaly, when the naive 
 continuum 
definition of $Q_5$ is used \cite{KarSm81}. 
In the euclidean formalism, the doublers can be removed by adding a 
higher derivative term to the action: the Wilson term, proportional to 
a parameter $r$ \cite{Wil75}. 
Another possibility is to interpret the doublers 
as physical particles, i.e. the lattice action describes more than 
one flavour. This is the idea behind the staggered fermion method 
\cite{stagg}.

In this paper, we add a Wilson term in space, proportional to $r_1$,
to suppress the space doublers.
We interpret the doublers in time as physical particles. 
Hence the lattice action describes two flavours. In \cite{AaSm98b} we 
discuss in detail fermions on a lattice in real-time, 
and we construct the Hilbert space and the transfer matrix, 
associated to the lattice fermion path integral, which clarifies further the 
physical interpretation of the lattice system.

The fermion part of the lattice action is given by
\be
\label{eqactionqed}
S_F = -\sum_x \left[ \bar\psi(\Dslash +W)\psi 
+\half G\phi^*\psi^T\CC^\dagger\psi - \half 
G\phi\bar\psi\CC\bar\psi^T\right].\ee 
The Dirac operator is
\[
\Dslash = \sum_\mu\gm^\mu \bar D_\mu,\;\;\;\; 
\bar D_\mu  =  \half \left(D_\mu + D_\mu'\right),\]
which is explicitly (we recall that the fermions have charge $q=\half$)  
\[
\bar D_{\mu xy} = 
\frac{1}{2a_\mu}\left[
\delta_{x+\hat\mu a_\mu,y} U^q_{\mu}(x) - \delta_{x,y+\hat\mu a_\mu} 
U^{q\dagger}_{\mu}(y)\right],\;\;\;\;
 U_{\mu}^q(x) = e^{-iqa_\mu A_{\mu}(x)}.
\]
Note that $\bar D_\mu^\dagger = - \bar D_\mu$, and the (anti)symmetric parts 
are given by 
$\bar D_\mu^S = \mbox{Im}\bar D_\mu, \bar D_\mu^A = 
\mbox{Re}\bar D_\mu$. 
The Wilson term (in space and time, with parameters $r_\mu=(r_0,r_1)$), 
is 
\bean W &=& \sum_\mu W^\mu,\;\;\;\; W^\mu = -\half a_\mu r_\mu D'_\mu 
D^\mu,\\
 W_{xy}^\mu &=& -\frac{r_\mu}{2a_\mu}
\left[ -2\delta_{x,y} + 
\delta_{x+\hat\mu a_\mu,y} U^q_{\mu}(x) + \delta_{x,y+\hat\mu a_\mu} 
U^{q\dagger}_{\mu}(y)\right],\eean
with  $W^\dagger = W$, and the (anti)symmetric parts  
 $W^S = \mbox{Re}W, W^A = \mbox{Im}W$. 
As indicated above, we use $r_0=0$ throughout this paper.
The numerical results presented further on, are obtained using  
$r_1=1$.\footnote{ In the original model of the previous section, 
the Wilson term corresponds to a Majorana-Wilson term 
\[ S_{F,\rm W} =  - \sum_x \left[
\psi^T {\cal C}^\dagger P_L W\psi - \bar\psi P_R {\cal 
C}W\bar\psi^T\right].
\]
}

Because of the Majorana coupling between the scalar and the fermion field, 
it is convenient to use a Majorana formulation with 
real gamma matrices and a  
four-component real Majorana field $\Psi$. It is constructed from the real 
and imaginary parts of $\psi$, 
\[   \psi = \frac{1}{\sqrt{2}}(\Psi_1-i\Psi_2), \;\;\;\;
   \psi^\dagger = \frac{1}{\sqrt{2}}(\Psi_1^T+i\Psi_2^T), \;\;\;\;
\Psi = 
 \left(\begin{array}{c} \Psi_1\\ \Psi_2 \end{array}\right).
\]
We also introduce Pauli matrices, written as $\rho_{1,2,3}$, that act on 
the 1,2 components of $\Psi$.

In terms of $\Psi$ and the $\rho_i$ matrices, the fermion part of the 
action can be written as 
\bean
S_F &=& -\sum_x \half \Psi^T\beta\left[ 
\Dslash^A + \rho_2\, \Dslash^S + W^S+\rho_2W^A + 
\frac{G}{\sqrt{2}}(\phi_1\rho_3+\phi_2\rho_1)\right]
\Psi\\
&\equiv& -\sum_x  \half \Psi^T\beta \left[ \calDslash +{\cal W} 
+G\Phi\right]\Psi.
\eean
Here we defined the covariant derivative and the Wilson term on a 
Majorana field as 
\[ {\cal D}_\mu = \bar D_\mu|_{q\to q\rho_2},\;\;\;\;
 {\cal W} = W|_{q\to q\rho_2},\]
i.e.
\[ U_{\mu}^q(x)\to {\cal U}^q_{\mu}(x) = e^{-iq\rho_2 a_\mu
A_{\mu}(x)};\] 
$q\rho_2$ is the Majorana charge matrix.
For the Yukawa part we used
the explicit representation $\CC =\beta$, 
\[ \phi =\frac{1}{\sqrt{2}}(\phi_1 - i\phi_2),\] 
and we introduced a matrix field notation
\[ \Phi =\frac{1}{\sqrt{2}}(\phi_1\rho_3+ \phi_2\rho_1).\] 
The lattice field equation (in the temporal gauge) reads
\be
\label{eqdiraceq}
 \half i(\partial_0+\partial'_0)\Psi = \HD\Psi,\ee
where the Majorana Dirac hamiltonian is given by ($\al^1=-\gm^0\gm^1$) 
\be \HD = -i\alpha^1 {\cal D}_1 + \beta ({\cal W}^1 + G\Ph).\ee
It is hermitian and antisymmetric ($\HD^\dagger= \HD = -\HD^T$). 

 The lattice fermion doublers in space are removed by the Wilson term 
in space. 
We now describe how we deal with fermion doubling
due to the `naive' time discretization ($r_0=0$).
Reverting to ordinary Dirac notation, the free fermion lattice propagator
is given by
\be
S(p,\om) = \frac{m_{Fp}  - i\gm_1 a^{-1}\sin ap + i\gm^0 a_0^{-1}\sin a_0\om}
{m_{Fp}^2 + a^{-2}\sin^2 ap - a_0^{-2}\sin^2 a_0\om}.
\ee
 The fermion mass term $m_{Fp}= m_F + m_p$ is the sum of the physical 
continuum mass term $m_F$ and a term due to the higher derivative in 
space, the Wilson `mass term', $m_p=a^{-1} r_1 (1-\cos ap)$. 
This propagator 
takes the usual continuum form for $a,a_0 \to 0$.
However, the propagator has also a pole near $a_0\om = \pm \pi$, corresponding
to the rapid time variation $(-1)^{t/a_0}$. As shown
in \cite{KarSm81}, such poles have to be interpreted as particles, the
doublers. Their physical frequency $p^0$ is related to $\om$
by $a_0\om = \pi + a_0 p^0$, such that we recover the continuum propagator
in the limit $a,a_0 \to 0$, up to the unitary 
transformation $i\bt\gm_5 = -\gm_1$,
\be
S(p,\frac{\pi}{a_0} + p^0) \to \gm_1\left(
\frac{m_F-i\gm_1 p + i\gm^0 p^0}{m_F^2 + p^2 - p_0^2}\right)\, \gm_1.
\ee
Hence, the lattice fermion field $\ps$ describes
{\em two} physical fields in the continuum limit, 
 which we label with flavour indices $u,d$, and denote with 
$\ch_u$ and $\ch_d$,
\be
\ps(x,t) \to \ch_u(x,t) - (-1)^{t/a_0} \gm_1 \ch_d(x,t).
\ee
In Majorana notation,
$\ch = (X_1 -i X_2)/\sqrt{2}$,
this takes the form
\be
\Psi(x,t) \to  X_u(x,t) - (-1)^{t/a_0} \gm_1\rh_2 X_d(x,t).
\label{PsiX}
\ee
Substitution in the lattice action displays its physical content in
the continuum limit 
\be
S_F \to -\int d^2 x\, \half X^T \bt\left[\gm^{\mu} D_{\mu}
+ G\Phi\ta_3\right] X.
\ee
Here we combined $X_{u,d}$ into a flavour doublet $X$ on which Pauli
matrices $\tau_{1,2,3}$ act. Cross terms $\propto (-1)^{t/a_0}$ do not 
contribute in the continuum limit.
A useful relation is 
\be \label{eqHG}
\gm_1\rho_2 \HD(+G)\gm_1\rho_2 = -\HD(-G).
\ee

The Majorana Yukawa coupling happens to break the SU(2) flavour
symmetry of the continuum limit of the $G=0$ theory.  Experience with
euclidean lattice fermions suggests that it could be avoided, if so desired,
and that we can obtain a wide class of target continuum theories
in a more sophisticated staggered fermion formulation. 

It is convenient to formulate the back reaction of the fermions on the
Bose fields 
 (in sections \ref{secaction} and beyond)
in terms of initial expectation values of fermion operators.
This calls for an extension of (\ref{PsiX}) to an operator equation.
Since the continuum fields $X_u, X_d$ are supposed to be slowly varying in 
lattice units, they are somewhat spread out over the lattice. 
A minimal spreading in time is over two time slices, which is also
used in derivations of the quantum mechanical Hilbert space from
the lattice path integral \cite{Sm91}. Labeling pairs of time slices by
an integer $k$, we may write without `overcounting', for even $t/a_0=2k$,
\bea
\Psi(x,t) &=&  X_u(x,t) -\gm_1\rh_2 X_d(x,t),\nonumber\\
\Psi(x,t+a_0) &=& X_u(x,t) + \gm_1\rh_2 X_d(x,t),
\label{eqlatphys}
\eea
This relation suggests that it is also useful to combine $\Psi(x,t)$ and 
$\Psi(x,t+a_0)$ in a doublet,
\be
\label{eqPsidoublet}
\Xi(x,k) = \frac{1}{\sqrt{2}}\left(
\begin{array}{c}
\Psi(x,t+a_0) \\
\Psi(x,t) 
\end{array}
\right),
\ee
so that (\ref{eqlatphys}) can be written concisely as
\be
\label{eqXRX}
\Xi = RX,
\ee
with
\be
\label{eqR}
R=\frac{1}{\sqrt{2}}\left(
\begin{array}{cc}
1 & \gm_1\rho_2  \\
1 & -\gm_1\rho_2
\end{array}
\right),
\;\;\;\;
RR^\dagger=1,\;\;\;\; RR^T=\tau_1.
\ee
Eqs. (\ref{eqlatphys}) (or equivalently (\ref{eqXRX})) can now be viewed as 
relations
between operators.  In \cite{AaSm98b} we will give more details on the 
Hilbert space aspects of fermion doubling in real time. 
For now we assume that the physical Majorana fields $X$ correspond to
hermitian operators satisfying the anticommutation relations
\be
\label{eqXX}
\{X(x,t),X^T(y,t)\} = \dl_{x,y}.
\ee
Note that as operators, the $\Psi$'s are not hermitian because of the 
imaginary $\rh_2$. Typically, it 
is the Grassmann variables 
$\ps(x,t+a_0)$ and $\ps^+(x,t)$ which correspond to hermitian conjugate 
operators in Hilbert space \cite{Sm91,AaSm98b}.

We can expand the fields $X$ in eigenspinors of the Dirac hamiltonian in 
flavour space, 
\be
\label{eqHDfl}
\HD_{\rm fl} = -i\alpha^1 {\cal D}_1 + \beta ({\cal W}^1 +\tau_3 G\Ph),
\ee
and identify the coefficients in the expansion with creation 
and annihilation operators. For example, let $\tilde U_{\al u}(x), 
\tilde U_{\al d}(x)$
be a complete orthonormal set of positive 
energy\footnote{The negative energy solutions are given by $\tilde U^*$.} 
eigenspinors of $\HD_{\rm fl}$ at $t=0$.
Here $\al$ labels the set (which typically contains momentum), 
and $u,d$ indicates the flavour. 
We then expand 
\be
\label{eqexpXu}
X_{u}(x, 0) = \sum_{\al} \left[ 
b_{\al u}\tilde U_{\al u}(x) + b^\dagger_{\al u}\tilde U^*_{\al u}(x)\right], 
\ee
and similar for $d$.
The creation and annihilation operators obey the usual 
anticommutation relations 
\be
\label{eqbb}
 \{b^\dagger_{\al f}, b_{\al' f'}\} = 
\delta_{\al\al'}\delta_{ff'}, \;\;\;\;f=u,d.
\ee
We can now express the lattice field $\Psi$ at time slices $t/a_0=0,1$ in
terms of the eigenspinors and the operators, if we combine (\ref{eqexpXu}) 
with (\ref{eqlatphys}) at $k=0$.  This gives 
\bean \Psi(x,0) &=& \sum_\al \Big[ 
b_{\al u}\tilde U_{\al u}(x) + b^\dagger_{\al u}\tilde U^*_{\al u}(x) \\
&&\;\;\;\;\;\;\;\;
- b_{\al d}\gm^1\rho_2 \tilde U_{\al d}(x) - 
b^\dagger_{\al d}\gm^1\rho_2 \tilde U^*_{\al d}(x)
\Big],\\ 
\Psi(x,a_0) &=& 
\sum_\al \Big[ 
b_{\al u}\tilde U_{\al u}(x) + b^\dagger_{\al u}\tilde U^*_{\al u}(x)\\
&&\;\;\;\;\;\;\;\;
+ b_{\al d}\gm^1\rho_2 \tilde U_{\al d}(x) + 
b^\dagger_{\al d}\gm^1\rho_2 \tilde U^*_{\al d}(x)
\Big]. 
\eean 
This relation can be extended to arbitrary times by writing
\be 
\label{eqexpPsi}
\Psi(x,t) = 
\sum_\al \left[ b_{\al u}U_{\al u}(x, t) +
b^\dagger_{\al u}U^*_{\al u}(x, t) - b_{\al d}U_{\al d}(x,t) +
b^\dagger_{\al d} U^*_{\al d}(x,t)\right], \ee 
where
$U_{\al u}(x,t), U_{\al d}(x,t)$ are solutions of the lattice Dirac 
equation (\ref{eqdiraceq}), with the initial conditions
\bea
\nonumber
U_{\al u}(x,0) &=& U_{\al u}(x,a_0) = \tilde U_{\al u}(x),\\
\label{eqinitialU}
U_{\al d}(x,0) &=& -U_{\al d}(x,a_0) = \gm^1\rho_2\tilde U_{\al d}(x).
\eea
We shall refer to the $U(x,t)$'s as the mode functions.
We show in \cite{AaSm98b} that solving the Dirac equation for the mode 
functions $U$ is equivalent to time evolution in Hilbert space by 
means of the transfer operator.
Note that (\ref{eqinitialU}) shows that the ordinary modes start smoothly 
in time, while the doubler modes start like $(-1)^{t/a_0}$.

The mode functions remain orthonormal under the discrete time evolution
in the appropriate inner product. In terms of
\be
V_{\al f}(k) = \frac{1}{\sqrt{2}}\left(\begin{array}{c}U_{\al f}(t+a_0)\\
U_{\al f}(t)\end{array}\right),
\;\;\;\; t=2ka_0,\;\;\;\; f=u,d,
\ee
where we suppress the space dependence,
the time evolution can be written as
\bean
V_{\al f}(k+1) &=& M(k)\, V_{\al f}(k),\\
M(k) &=& \left(\begin{array}{cc} 
1-4a_0^2 \HD_{2k+2} \HD_{2k+1} & -i2 a_0 \HD_{2k+2}\\ -i2a_0 \HD_{2k+1} &1
\end{array}\right),
\eean
where the subscript $k$ indicates the time dependence of the Bose fields in 
the 
Majorana Dirac hamiltonian. The matrix $M$ is unitary in `the $\tau_1$ 
inner product',
$M^{\dagger} \tau_1 M = \tau_1$ and it follows that the norm of the $V's$ 
is conserved in this inner product. The orthonormality relations follow
from the initial conditions:  
$V_{\al f}^{\dagger}(0) \tau_1 V_{\al'f'}(0) = \dl_{\al\al'}\, 
(\tau_3)_{ff'}$.
Similarly the anticommutation relations of 
the lattice Majorana field are preserved: 
$\{\Xi(x,k),\Xi^T(y,k)\} = \tau_1\dl_{x,y}$, 
which follows directly from (\ref{eqXRX}-\ref{eqXX}).
These conservation relations have the same meaning as the Wronskian 
condition \cite{cosmo}-\cite{nonequi}.
They imply that the basic anticommutation relations, (\ref{eqXX}) and 
(\ref{eqbb}), remain valid also for times larger than zero.

To summarize this section, we have formulated the model of section 
\ref{seccontinuum} on a spacetime lattice. A large part has been devoted 
to the physical interpretation of the lattice fermion field. It 
represents two physical fermion fields in the continuum. In relation with 
the dynamics to be treated in sections \ref{secaction} and beyond, the 
important result of this section is the mode function expansion of the 
lattice field (\ref{eqexpPsi}) in terms of time independent operators 
(\ref{eqbb}) and spinor mode functions. These mode functions are 
solutions of the Dirac equation (\ref{eqdiraceq}) with initial conditions 
(\ref{eqinitialU}).

\section{Symmetries, currents and observables}
\label{secsymmetries}
\setcounter{equation}{0}

 We discuss several exact symmetries of the lattice action, since 
these lead to exactly conserved currents for finite lattice spacing.
Furthermore, we construct other observables that are related to 
symmetries of the continuum theory.

The local gauge symmetry under which the action is invariant, acts on the 
fields as
\[
\begin{array}{cccccc}
A_{\mu}(x) &\to & A_{\mu}(x) -\partial_\mu \xi(x), &
U_{\mu}(x) &\to & e^{-i\xi(x)} U_{\mu}(x) e^{i\xi(x+\hat a_\mu)},\\ 
\phi(x) &\to& e^{-i\xi(x)}\phi(x),& 
\Phi(x) &\to&  e^{-iq\rho_2\xi(x)}\Phi(x) e^{iq\rho_2\xi(x)},\\
\psi(x) &\to& e^{-iq\xi(x)}\psi(x),& \Psi(x) &\to& 
e^{-iq\rho_2\xi(x)}\Psi(x). \end{array}
\]
We indicated how the fields transform 
for the different ways in which they are written.

The global gauge symmetry leads to the conserved current $j^\mu(x) = 
j^\mu_{h}(x) + j^\mu_{f}(x)$, with the scalar part
\be
\label{eqhiggscurrent}
 j^\mu_{h}(x) =  i(D^\mu\phi(x))^*\phi(x) 
-i\phi^*(x)D^\mu\phi(x),\ee
and the fermion part (in Dirac and in Majorana notation)
\bea
\label{eqfermionc}
 j^\mu_{f}(x) &=& iq\left[
\bar\psi(x+\hat a_\mu) P_+^\mu U^{q\dagger}_{\mu}(x)\psi(x)
-\bar\psi(x) P_-^\mu U^q_{\mu}(x)\psi(x+\hat a_\mu)
\right]\\
\nonumber
&=& 
\frac{i}{2}\left[ 
\Psi^T(x+\hat a_\mu) \beta P_+^\mu q\rho_2 {\cal U}_{\mu}^{q\dagger}(x)
\Psi(x)
-
\Psi^T(x) \beta P_-^\mu q\rho_2 {\cal U}_{\mu}^q(x)
\Psi(x+\hat a_\mu)\right].
\eea
We use the notation, familiar from euclidean lattice fermions, $P_\pm^\mu = 
(r_\mu \pm \gm^\mu)/2$ (recall that $r_0=0$). It is easy to check what 
this current represents in terms of the 
flavour fields $X_u, X_d$, using the relations (\ref{eqlatphys}). For 
example, the charge density at even $t/a_0$ is
\bean
 j^0_f(x,t) &=& 
\frac{1}{4}\left[ 
\Psi^T(x,t+a_0) q\rho_2 \Psi(x,t)
+
\Psi^T(x,t) q\rho_2 \Psi(x,t+a_0)\right]\\
&=&
\frac{1}{2}\left[ 
X^T_u(x,t) q\rho_2 X_u(x,t)
+
X^T_d(x,t) q\rho_2 X_d(x,t)\right],
\eean 
which is precisely the charge density due to two Majorana fermions with 
charge $q$, as expected.

There is another global symmetry, due to the fact that the model is 
defined on a lattice. It is inspired by the staggered fermion 
formalism. It is 
\[ \Psi(x,t) \to \exp\left[i\om\gm^1(-1)^{t/a_0}\right] \Psi(x,t).\]
The associated charge density is
\[
 j^0_{\rm fl}(x,t) = 
\frac{(-1)^{t/a_0}}{4}\left[\Psi^T(x,t)\gm^1\Psi(x,t+a_0) - 
\Psi^T(x,t+a_0)\gm^1\Psi(x,t)\right].\]
The physical interpretation becomes clear in terms of the flavour 
fields. At even $t/a_0$, 
\bean
 j^0_{\rm fl}(x,t) &=& 
\frac{1}{2}\left[X^T_{u}(x,t)\rho_2X_{d}(x,t) + 
X^T_{d}(x,t)\rho_2X_{u}(x,t)\right] \\
&=&
\frac{1}{2}X^T(x,t)\rho_2\tau_1X(x,t),
\eean
where the last line is written compactly in flavour space.
It is the first component of the flavour triplet current.

Other observables, which are not directly related  to a symmetry of the 
lattice action, are constructed with the continuum 
limit as a guidance. The energy density of the fermions is found to be
\be
\label{eqfermionenergy}
H_f(x,t) = \frac{1}{4}\left[ 
\Psi^T(x,t)\HD(x,t)\Psi(x,t) + (t\to t+a_0)
\right]. \ee
In terms of the flavour fields at even $t/a_0$, 
 using again (\ref{eqlatphys}),
this is the expected energy density 
\[
H_f(x,t) = \frac{1}{2}
X^T(x,t)\HD_{\rm fl}(x,t)X(x,t),
\]
where $\HD_{\rm fl}$ was defined in (\ref{eqHDfl}), and we used (\ref{eqHG}).
The anomalous axial vector current which satisfies the anomaly 
equation is the flavour singlet current, i.e. where the contributions 
from both flavours add,
\bean
 j^0_5(x,t) &=&
\frac{1}{2}X^T(x,t) \gm_5\rho_2X(x,t)\\
&=&
\frac{1}{2}\left[ 
X^T_{u}(x,t) \gm_5\rho_2 X_{u}(x,t)
+
X^T_{d}(x,t) \gm_5\rho_2 X_{d}(x,t)\right].
\eean
It is straightforward to verify for even $t/a_0$, that in terms of the 
lattice field the axial charge density is obtained by taking  
\bean
 j^0_5(x,t) &=& 
\frac{1}{4}\left[ 
\Psi^T(x,t) \gm_5\rho_2 \Psi(x,t)
+ (t\to t+a_0)\right].
\eean 
We will divide, in the rest of the paper, the axial charge (density) by a 
factor of two to keep the anomaly equation, given in section 
\ref{seccontinuum}: $Q_5(t) - Q_5(0) = C(t)-C(0)$.

The current associated 
with the global symmetry 
$\psi  \to  \exp\,(-i\om\gm_5)\,\psi$
of the continuum action, is given by
\bean
j^\mu_{{\rm fl}5}(x) &=& \frac{i}{2}\left[
\bar\psi(x) \gm^\mu \gm_5 U^q_{\mu}(x)\psi(x+\hat a_\mu)
+
\bar\psi(x+\hat a_\mu)\gm^\mu\gm_5
U^{q\dagger}_{\mu}(x)\psi(x)
\right],\\
&=&
\frac{i}{4}\left[ \Psi^T(x) \beta\gm^\mu\gm_5\rho_2
{\cal U}^q_{\mu}(x)\Psi(x+\hat a_\mu)
+
\Psi^T(x+\hat a_\mu) \beta\gm^\mu\gm_5
{\cal U}_{\mu}^{q\dagger}(x)\Psi(x)\right].
\eean 
The continuum interpretation of this current is not that it represents the 
anomalous axial vector current, but rather the third component of the 
flavour {\em 
triplet} axial vector current. For example, the charge density at even 
$t/a_0$ is 
\bean
 j^0_{{\rm fl}5}(x,t) &=& 
\frac{1}{4}\left[ 
\Psi^T(x,t+a_0) \gm_5\rho_2 \Psi(x,t)
+
\Psi^T(x,t) \gm_5\rho_2 \Psi(x,t+a_0)\right]\\
&=&
\frac{1}{2}\left[ 
X^T_{u}(x, t) \gm_5\rho_2 X_{u}(x, t)
-
X^T_{d}(x,t) \gm_5\rho_2 X_{d}(x,t)\right]\\
&=&
\frac{1}{2}X^T(x,t) \gm_5\rho_2\tau_3 X(x,t).
\eean 
Indeed, this density represent the difference between the two 
axial charge densities instead of the sum. This is known in the 
naive fermion formalism as well \cite{KarSm81}. 
The current is not exactly conserved because the global symmetry is 
broken explicitly on the lattice by the Wilson term in space. 

After this lengthy discussion of the model and its lattice version, we 
come to the effective action and equations of motion in the 
following section.

\section{Large $N_f$ effective action}
\label{secaction}
\setcounter{equation}{0}

In this section we derive the effective action by solving the path 
integral to  leading order in the large $N_f$ approximation. Since this 
can already be found in great detail in the literature \cite{CoMo94, Boya96}, 
we only show the essential steps.
The effective equations of motion follow directly from the effective action.

To avoid a notational jam, it is convenient to denote the Bose fields 
collectively with $\varphi =(\phi, \phi^*, A_\mu)$. We also denote 
the corresponding sources as $J = (J^*, J, J^\mu)$. 
We use a compact notation, such that e.g. $J\cdot\varphi = 
\sum J(x)\varphi(x)$, where the sum is over all indices, including 
space-time. 

\begin{figure}
\centerline{\psfig{figure=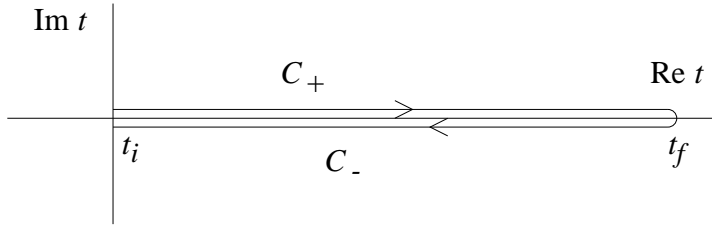,height=3.0cm}}
\caption{Keldysh contour in the complex time plane.}
\label{figkeldysh}
\end{figure}

The partition function we are interested in is
\be 
Z[J] = \Tr\, \rho T_C \exp\, iJ\cdot\varphi.
\ee
$T_C$ indicates that the 
exponent is time ordered along the Keldysh 
contour in the complex time plane, which is shown in figure 
\ref{figkeldysh}. 
We take the initial density matrix $\rho$ such that the fermion field has 
zero mean value.
Matrix elements of the initial density matrix are written as
\[ \bra \varphi_a, \Psi_a|\rho| \varphi_b, \Psi_b\ket  = 
\rho(\varphi_a, \Psi_a; \varphi_b, \Psi_b).\]
The corresponding path integral expression reads
\be
Z[J] = \int {\cal D}\varphi_a
{\cal D}\varphi_b
{\cal D}\Psi_a
{\cal D}\Psi_b\,
\rho(\varphi_a, \Psi_a; \varphi_b, \Psi_b)
\int_C
{\cal D}\varphi
{\cal D}\Psi
\exp (iS+iJ\cdot\varphi),\ee
with $S=S_B+S_F$, $S_B=S_G +S_H$. The 
{\scriptsize $C$} again denotes that the whole expression is defined  
along the Keldysh contour. The initial values of the fields and 
correlations are determined by the density matrix.

Let's for a moment consider the part of the partition function that 
involves the fermion field. 
Formally, we can perform the integration over the fermion field, and we 
denote the result as 
\be
\label{eqlogD}
\half \log D[\varphi]
\equiv
\log \int {\cal D}\Psi_a
{\cal D}\Psi_b\,
\rho(\varphi_a, \Psi_a; \varphi_b, \Psi_b)
\int_C
{\cal D}\Psi\,
\exp iS_F[\varphi].
\ee
$D[\varphi]$ depends on the Bose fields $\varphi$ along the contour, 
including the initial Bose fields $\varphi_a, \varphi_b$.
Note that if the density matrix is quadratic in $\Psi$, $D$ represents 
the resulting determinant.
We would like to remark that it is not necessary to have a gaussian 
density matrix for $\Psi$ to perform the integration. Furthermore,
we do not need to specify any property of 
the density matrix regarding its Bose field dependence.

To arrive at a large $N_f$ expansion, we duplicate the field $\Psi$ $N_f$ 
times, and demand that all fields $\Psi^i$ are initialized with the same 
density matrix, i.e.
\[
\Psi \to \Psi^i, \;\;\;\; 
\rho(\varphi_a, \Psi_a; \varphi_b, \Psi_b) \to
\prod_i \rho(\varphi_a, \Psi_a^i; \varphi_b, \Psi_b^i),\;\;\;\;
i=1,\ldots, N_f.\]
If we then integrate out the fermion fields, we find 
\[
Z[J] = \int {\cal D}\varphi_a
{\cal D}\varphi_b
\int_C {\cal D}\varphi\,
\exp\,i \left(S_B -i\half N_f\log D
+ J\cdot\varphi\right).
\]
To proceed, it is common use to rescale the parameters, and the Bose 
fields and sources such that the exponent in the path integral becomes 
proportional to $N_f$. In our model, this rescaling is explicitly 
\bean
\begin{array}{ccccccccc}
\phi &=&  \phi'\sqrt{N_f}, &  v &=&  v'\sqrt{N_f},& J   &=& 
J'\sqrt{N_f},\\
 e   &=&   e'/\sqrt{N_f}, &  G &=& G'/\sqrt{N_f}, & \lm &=& 
\lm'/N_f. \end{array}
\eean
Note that this rescaling implies that 
the original, unscaled theory is weakly coupled
when $N_f$ is large, and that the initial 
conditions for the unscaled $\phi$ and $A_\mu/e$
have to be such that they are of order $\sqrt{N_f}$.

In terms of the primed variables, the partition function is (up to 
normalization) \be
\label{eqsaddle}
Z[J'] = 
\int {\cal D}\varphi'_a
{\cal D}\varphi'_b
\int_C {\cal D}\varphi'\,
\exp\, iN_f\left(S'_B -i\half\log D' + J'\cdot\varphi'\right).
\ee
which suggests
a saddle point expansion for large $N_f$.
The field configurations that determine the stationary phase, 
$\varphi_s$, are found by extremizing the exponent of (\ref{eqsaddle}). 

A Legendre transformation from the generating functional $\log Z$ 
depending on the sources, to the effective action depending on the so-called 
classical or mean fields, 
\[ 
Z[J'] = \exp\,iN_fW[J'],\;\;\;\;\;
\bar\varphi = \frac{\partial W[J']}{\partial J'}, \;\;\;\;\; 
\Gm[\bar\varphi] = W[J'] - J'\cdot\bar\varphi,
 \]
yields the effective action, 
which is to leading order in large $N_f$, 
\[ \Gamma[\bar\varphi] = S_B[\bar\varphi] -\frac{i}{2} \log 
D[\bar\varphi].\]
The effective equations of motion follow from
\[  \frac{\partial \Gm[\bar\varphi]}{\partial\bar\varphi} = -J'.\]

The mean fields $\bar\varphi$ are defined on the complete Keldysh 
contour. In general, the sources $J$ are different on the 
upper and lower part of the contour (which is needed to be able
to construct correlation functions of arbitrary observables). 
This implies that also the mean fields are different.  
However, having derived the equations of motion and expressions
for the observables, we can set the upper and lower sources equal,
which implies that the upper and lower mean fields are also 
the same. The initial conditions are in principle to
be derived from the density matrix. 
This can be analyzed conveniently if it has a gaussian form.
To leading order $\varphi_s = \bar\varphi$ \cite{CoMo94}.

Differentiating $\log D$ at $t>t_i$ gives the 
back reaction of the fermion field in the background of the 
Bose fields, at time $t$. 
We use a bracket notation to denote the result 
\[
\frac{1}{2}\frac{\partial}{\partial \bar\varphi(x,t)}\log D
\equiv i\bra \frac{\partial S_F}{\partial \varphi(x,t)}\ket. \]
The initial conditions for the fermion field are also determined 
by (\ref{eqlogD}). However, we find that is more transparent to use the 
operator formalism. We discuss the fermion initial conditions in the following 
section.

To conclude this section,
we would like to remark that in this approximation scheme the fermion 
fields are integrated out exactly. The large $N_f$ approximation is only 
used to justify a saddle point expansion of the resulting effective 
bosonic action. The leading order result in this expansion is that the Bose 
fields are treated classically, and the fermions show up in a one loop 
contribution, but with the exact propagator. 
In the following, we drop the bars from the mean fields.

\section{Effective equations of motion}
\label{secequations}
\setcounter{equation}{0}

We are now in a position to give the effective equations of motion.
Using the results from the previous section, we find  
the bosonic field equations (in the temporal gauge)
\bea
\label{eqAeq}
\partial'_0\partial_0 A_{1}(x,t) \!\!&=&\!\! e^2 \left[j_{h}^1(x,t) + \bra 
j_{f}^1(x,t)\ket\right],\\ 
\label{eqgausslaw}
\partial'_1\partial_0 A_{1}(x,t) \!\!&=&\!\! -e^2 \left[j_{h}^0(x,t) + \bra 
j_{f}^0(x,t)\ket\right],\\ 
\label{eqhiggs}
\partial'_0\partial_0 \phi(x,t) \!\!&=&\!\! D'_1 D_1\phi(x,t) - 2\lambda 
\left[|\phi(x,t)|^2 -  v^2_B/2\right]\phi(x,t) + G\bra F(x,t)\ket.
\eea
The scalar and fermion contribution to the current, $j^\mu_h$ and $j_f^\mu$ 
respectively,  are given by (\ref{eqhiggscurrent}) and (\ref{eqfermionc}).
The force on the Higgs field due to the fermions is defined by 
\be
\label{eqFfh}
F(x,t) \equiv \frac{1}{G}\frac{\partial S_F}{\partial \phi^*(x,t)}  =
i\frac{1}{4}\Psi^T(x,t)\beta(\rho_1+i\rho_3)\Psi(x,t).
\ee
Eq. (\ref{eqgausslaw}) is Gauss' law. It implies 
(on a periodic lattice) that the total charge is 
zero, 
\be
\label{eqgaussglobal}
 Q(t) = Q_h(t)+Q_f(t) = \sum_{x} \left(j^0_{h}(x,t) + \bra 
j^0_{f}(x,t)\ket\right) =0.
\ee
In (\ref{eqhiggs}), we used a subscript {\scriptsize $B$} to indicate 
that $v^2_B$ is a bare parameter. We'll discuss renormalization in 
section \ref{secrenorm}.

Following \cite{pairpro}-\cite{nonequi} for the case of homogeneous 
classical fields, 
we calculate the back reaction of the fermions on the Bose fields using 
a mode function expansion, introduced in section \ref{seclattice}. 
We expand the field $\Psi$ in a complete set of eigenspinors of the 
initial Dirac hamiltonian, with time-independent creation and annihilation 
operators as coefficients (i.e. (\ref{eqexpPsi}))
\be
\Psi(x,t) = 
 \sum_\al \left[ b_{\al u}U_{\al u}(x, t) +
b^\dagger_{\al u}U^*_{\al u}(x, t) - b_{\al d}U_{\al d}(x,t) +
b^\dagger_{\al d} U^*_{\al d}(x,t)\right].
\ee
Expectation values of the operators determine the initial quantum state 
of the fermion field.
Two typical choices are a vacuum and a thermal state. These are given by 
the following expectation values
\[ \bra b^\dagger_{\al f}b_{\al'f'}\ket = 
n_{\al f}\delta_{\al\al'}\delta_{ff'}
 \;\;\;\;
 \bra b_{\al f}b^\dagger_{\al'f'}\ket = 
(1-n_{\al f})\delta_{\al\al'}\delta_{ff'},\;\;\;\;f=u,d.\]
Here $n_{\al f}$ is the Fermi-Dirac distribution
\be
n_{\al f} = [\exp(E_{\al f}/T_{f, \rm in})+1]^{-1},
\label{FermiDirac}
\ee
with $E_{\al f}$ the eigenvalue of 
the Dirac hamiltonian for eigenspinor $U_{\al f}$, 
and $T_{f, \rm in}$ the initial temperature of the fermions.
We will often encounter
\be
\label{eqinit}
  \bra [b_{\al f},b_{\al f}^\dagger]\ket 
= 1-2n_{\al f}  \equiv \sg_{\al f}.
\ee
All fermion observables can now be expressed in terms of $\sg_{\al f}$ 
and the mode functions. 
Regarding the numerical results that are presented in this paper, we 
restrict ourselves to initial fermions in vacuum, i.e. $T_{f, \rm in}=0$, 
for which $n_{\al f}$ vanishes of course. 

The system is closed by giving the equation of motion for these mode 
functions. As explained in detail in section \ref{seclattice}, both  
$U_{\al u}$ and $U_{\al d}$ are solutions of the lattice Dirac equation 
(\ref{eqdiraceq}), 
\be
 \half i(\partial_0+\partial'_0)U_{\al f}(x,t) 
 = \HD\left[A_1(x,t), \phi(x,t)\right] U_{\al f}(x,t),\;\;\;\;f=u,d,
\ee 
with initial conditions (\ref{eqinitialU}).
The number of partial differential equations that has to be solved is 
quite large. 
Besides the Bose fields, there are $4N$ 
four-component complex mode 
functions, which make the numerical calculation scale as $N^3$ (for fixed 
$a_0/a$). 

To be explicit, we list the fermion field expectation values in terms of 
the mode functions. To avoid too lengthy formulas, we combine the two 
flavour mode functions in one, 
\[  U_\al = \left( \begin{array}{c} U_{\al u}\\ U_{\al d} 
\end{array}\right),\]
and use for the expectation values (\ref{eqinit}) the matrix
\[  \sg_\al = \left( \begin{array}{cc} \sg_{\al u} & 0 \\ 0 & -\sg_{\al d} 
\end{array}\right).\]
We stress however that the mode functions are still to be 
solved with the original Dirac hamiltonian (i.e. not the one in flavour 
space).

In the equations of motion we need the charge and current density, and 
the force the fermions exert on the scalar field. These are given by
\bean 
\bra j^0_f(x,t) \ket &=& -\frac{1}{4}\sum_{\al}\left[
U^\dagger_{\al}(x,t)q\rho_2 \sg_{\al}U_{\al}(x,t+a_0)
+U^\dagger_{\al}(x,t+a_0)q\rho_2\sg_{\al}U_{\al}(x,t)
\right],\\
\bra j^1_f(x,t) \ket &=& \frac{i}{2}\sum_{\al} \big[
U^\dagger_{\al}(x,t)\beta P_-^1q\rho_2\sg_{\al}
{\cal U}^q_{1}(x,t)U_{\al}(x+a,t)\\
&&\;\;\;\;
-U^\dagger_{\al}(x+a,t)\beta P_+^1q\rho_2\sg_{\al}
{\cal U}^{q\dagger}_{1}(x,t)
U_{\al}(x,t)
\big],\\
\bra F(x,t) \ket &=& -\frac{i}{4}\sum_{\al}\left[ 
U^\dagger_{\al}(x,t)\bt(\rho_1+i\rho_3) \sg_{\al}U_{\al}(x,t)\right].
\eean
Two observables we are interested in, are the anomalous axial 
charge and the conserved flavour charge. Their densities are
\bean
\bra j^0_5(x,t) \ket &=& -\frac{1}{4}\sum_{\al} \left[
U^\dagger_{\al}(x,t)\gm_5\rho_2\sg_{\al}U_{\al}(x,t)
+ (t\to t+a_0)
\right],\\
\bra j^0_{\rm fl}(x,t) \ket &=& 
\frac{(-1)^{t/a_0}}{4}\sum_{\al} \left[
U^\dagger_{\al}(x,t+a_0)\gm_1\sg_{\al}U_{\al}(x,t) -
U^\dagger_{\al}(x,t)\gm_1\sg_{\al}U_{\al}(x,t+a_0)
\right].
\eean
Finally, the energy density of the fermions is
\[ 
E_f(x,t) = -\frac{1}{4}\sum_{\al} \left[
U^\dagger_{\al}(x,t)\HD(x,t)\sg_{\al}U_{\al}(x,t)
+   (t\to t+a_0)
\right].
\]

All these formulas are valid for a complete set of mode functions that are
initially eigenspinors of an arbitrary Dirac hamiltonian, i.e. a 
Dirac hamiltonian with arbitrary Bose fields as background.
In the numerical treatment of the equations we 
present in the following sections, we use as initial background fields  
for the Dirac hamiltonian a vacuum configuration of Bose fields, i.e. 
$A_1=0, \phi=v_R/\sqrt{2}$, where $v_R$ is the renormalized vacuum 
expectation value. 
In terms of the density matrix, this means that 
we take no initial correlations between the Bose and the fermion fields, 
$\rho = \rho_B\otimes\rho_F$, and that $\rho_F$ is quadratic 
in $\Psi$. In Appendix \ref{appinitialf} we calculate the eigenvalues and 
-spinors in such a configuration, with the following results.

The energy eigenvalues can be summarized as
\[
E_{p\eta} = \sqrt{s_p^2 +m_{p\eta}^2},
\]
with 
\[
s_p = a^{-1}\sin pa, \;\;\;\; m_{p\eta} = m_p + \eta m_F, \;\;\;\; 
\eta =\pm.
\]
 Here $p$ is discrete and takes $N$ values, given by (\ref{eqp}). 
The total fermion mass is the sum or difference between the mass  
due to the Wilson term, $m_p = a^{-1}r_1(1-\cos pa)$, and the mass  
due to the Higgs vacuum expectation value, $m_F = Gv_R/\sqrt{2}$. 
The collective label is now explicit, $\al = (p, 
\eta)$, 
 and has $2N$ values. 
For these initial conditions, we find that the charge and current 
densities vanish initially,
\[ \bra j^0_f(x,0) \ket = \bra j^1_f(x,0) \ket =
 \bra j^0_5(x,0) \ket = \bra j^0_{\rm fl}(x,0) \ket = 0.\]
The initial fermion force on the scalar field is
\be
\label{eqFinit}
 \bra F(x,0) \ket = \bra F(x,a_0) \ket = 
\frac{1}{2L}\sum_{p\eta}\sg_{p\eta}\frac{\eta m_{p\eta}}{E_{p\eta}}.\ee
The initial energy density
\[ E_f(x,0) = -\frac{1}{L}\sum_{p\eta}\sg_{p\eta}E_{p\eta},\]
contains the zero temperature divergent part ($\sg_{p\eta}\to 1$). We 
renormalize this by subtracting the bare fermion energy density, 
\be
\label{eqbareenergy}
E_f^R(x,t) = E_f(x,t) - E_f^B,\;\;\;\; 
 E_f^B = -\frac{1}{L}\sum_{p\eta}E_{p\eta}.
\ee
The total energy of the system $E_{\rm tot}$ is the sum of the energy in the 
fermions and in the classical Bose fields, 
\be
E_{\rm tot}(t) = E_b(t)+E_f(t),\;\;\;\;
E_b(t) = \sum_x \left[ \half e^2 E^2(x,t) + |\pi(x,t)|^2\right] + V_{\rm 
pot}(t),
\ee
with
\bean 
&&e^2E(x,t) = \partial_0A_1(x,t), \;\;\;\;\pi(x,t) = 
\partial_0\phi(x,t),\\
&&V_{\rm pot}(t)  = \sum_x 
\left[ |D_1 \phi(x,t)|^2 + \lambda\left(|\phi(x,t)|^2 - 
v_B^2/2\right)^2\right].
\eean

Let us conclude this section with the following remarks.
When the initial state of the system is homogeneous, the system will remain 
homogeneous during the time evolution. The space dependence of the mode 
functions is given by plane waves for all times, and there is no mode 
mixing.  The equations of motion reduce to ordinary differential equations.
In the case of inhomogeneous initial conditions, the subsequent space 
dependence is determined by the equations. There is mode mixing, and 
the label $p$ looses its interpretation for $t>0$. 
Since the initial state of the fermions we use here, is choosen to 
be homogeneous, the inhomogeneity in  the initial conditions has to be put 
into the Bose fields.
This should be done in such a way that the total Higgs charge is zero, 
$Q_h(0)=0$ (since $Q_f(0)=0$) and that it does not violate the local Gauss' 
law (\ref{eqgausslaw}). In this one space dimensional model, this can be done 
in a straightforward manner.
Another possibility to introduce inhomogeneity would be to start not with a 
plane wave basis for the fermions, but with wave packets. This could be 
used to study scattering between fermions. This would correspond to an 
initial density matrix that is not quadratic in the fermion field.

Before diving into the full dynamics, we first analyse the response of 
the fermions to certain external Bose field configurations.

\section{Handmade sphaleron transitions}
\label{sechandmade}
\setcounter{equation}{0}

 In the previous section, we listed the effective equations of motion 
and the initial conditions for the fermion mode functions. Once the 
initial Bose fields are specified, the equations can be solved 
(numerically). In order to gain experience with the equations for the  
mode functions, we treat in this section the Bose fields as {\em 
external} fields. We focus on sphaleron transitions: 
the Bose fields start in a vacuum configuration, characterized by 
a certain integer Chern-Simons number $C$ and scalar field winding number 
$n$, go through a sphaleron, and end up in a vacuum where $C$ and $n$ have 
changed by one unit. 
This analysis is closely related to the spectral flow analysis, which we 
summarize in Appendix \ref{appspec}. However, when considering spectral 
flow, usually the Dirac hamiltonian is diagonalized for every external 
Bose field configuration \cite{Am94, Bock}. Here we use 
the real-time approach and solve the equations of motion for the mode 
functions in the presence of the Bose fields.
Such a calculation can be done analytically in the case of QED with a 
homogeneous gauge field \cite{Am83}. 
We do not have to discuss renormalization of the 
effective equations of motion at this stage, since the Bose fields are 
external, and the divergence is only present in the fermion back reaction.
The divergence in the fermion energy is renormalized as in 
(\ref{eqbareenergy}).
We stress again that the main purpose of this section is to test the 
equations for the mode functions.


The Bose configurations we use are the following.
A vacuum configuration with Higgs winding number $n$ is given by
\[  \phi_n = \frac{v}{\sqrt{2}}\exp \left(2\pi i n \frac{x}{L}\right)
=\frac{v}{\sqrt{2}}V_n.\]
Note that on the lattice $n$ is only defined modulo $N$, 
because $Nx/L= x/a$ is integer.
We use the gauge invariant definition 
of the winding number \cite{Ka98} on a periodic 
lattice, 
\bean 
\phi(x,t) &=& \rho(x,t) e^{i\theta(x,t)},\\
n(t) &=& \frac{1}{2\pi}\sum_x \Big(\left[\theta(x+a, t)-\theta(x,t) - 
A_1(x,t)\right]_\pi + A_1(x,t)\Big),\\
\left[f\right]_\pi &=& f \,(\mbox{mod } 2\pi) \in (-\pi, \pi].
\eean 
The corresponding gauge field configuration is
\[  A_1^n = \frac{2\pi n}{L}, \;\;\;\;C=-n.\]
The sphaleron configuration is given by
\[ \phi_s = -\frac{v}{\sqrt{2}}\tanh\left(\sqrt{\lambda/2} 
v(x-L/2)\right)\exp\left(i\pi\frac{x}{L}\right),
\;\;\;\;A_1=-\frac{\pi}{L}, \;\;\;\;C=\half.\]

We go from one vacuum to another through the sphaleron in a time $t_0$, 
with the sphaleron being halfway. $t_0$ sets the time scale.
Since we want to go through several vacua 
after each other, the configurations are taken to be ($t' = t/t_0$) 
\[
\phi(t') = f(t')\phi_{n_v} + (1-f(t'))\phi_sV_{n_s}, \;\;\;\;
A_1(t') = \frac{2\pi}{L}g(t'),\] 
where the winding number $n_v$ changes from $n_v$ to $n_v+1$ at $t'$ 
being half-integer, and $n_s$ from $n_s$ to $n_s+1$ at $t'$ being 
integer. At $t'=0$, $n_v=n_s=0$. The profile functions 
$f, g$ that determine the time-dependence obey the following conditions:
$f$ equals 1 at integer $t'$, and 0 at half-integer $t'$, 
and $g$ equals $t'$ at (half-)integer $t'$.
Furthermore, we require the time derivatives of $f, g$ to vanish at 
$t'$ being (half-) integer.\footnote{To be explicit, we use
$f(t')=\cos^2\pi t', g(t')=t' -(4\pi)^{-1}\sin
4\pi t'$.} This means that the configuration is `stationary' 
both on top of the sphaleron and in the groundstate.
Note that these configurations are inhomogeneous in general.

The parameters we use in this section are
\be
\label{eqparhandmade}
\begin{array}{llll}
N=32, & a_0/a=0.05, & eL=3.2, &  et_0 = 2,\\
 v^2=4, & \lm/e^2=0.25, & G/e=0.1, & 
\end{array}
\ee
except when 
indicated otherwise.
Let us shortly discuss the lattice spacing dependence. 
All the results we present are for a fixed physical size $eL$. 
Hence, increasing the number of lattice points $N$ reduces in fact the 
lattice spacing $a$. 
Furthermore, the time scale $et_0=2$ we take here is sufficiently large 
such that all physical time scales are large compared to lattice scales, but 
also not so large that the adiabatic approximation ($et_0\to \infty$) 
tells all.

\begin{figure}
\centerline{\psfig{figure=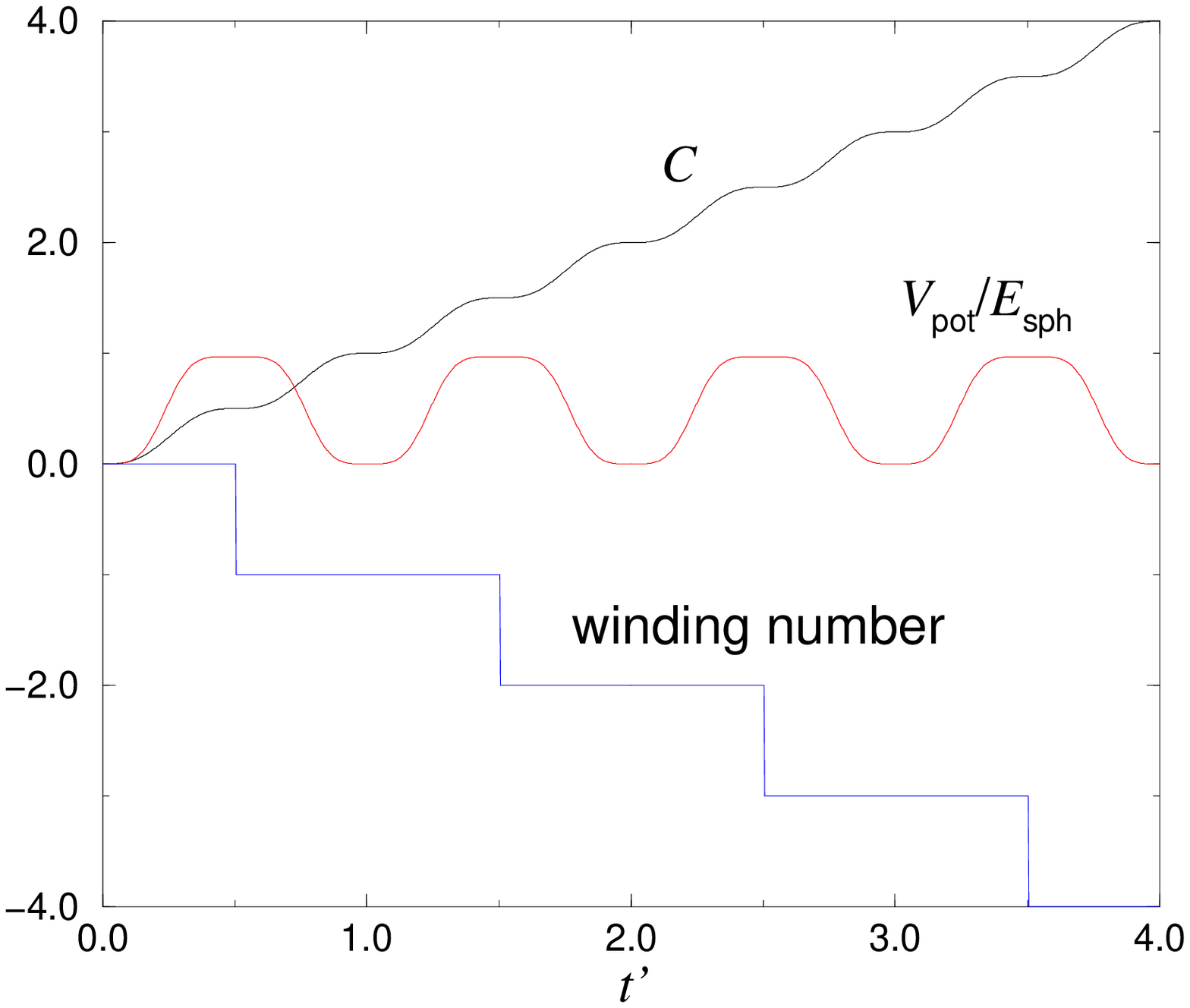,width=16.0cm}}
\caption{Characterization of the Bose fields configuration: 
the Chern-Simons number, the Higgs winding number and the 
potential energy in the Bose fields, normalized with the sphaleron 
energy, versus $t'$, for parameters (\ref{eqparhandmade}). }
 \label{fighandmade1}
\end{figure}

In figure \ref{fighandmade1}, we show a characterization of the Bose 
fields configuration, i.e. the Chern-Simons number, the winding number, 
and the potential energy in the Bose fields, normalized with the sphaleron 
energy. 
The potential energy in the Bose fields is the well-known periodic 
potential.

\begin{figure}
\centerline{\psfig{figure=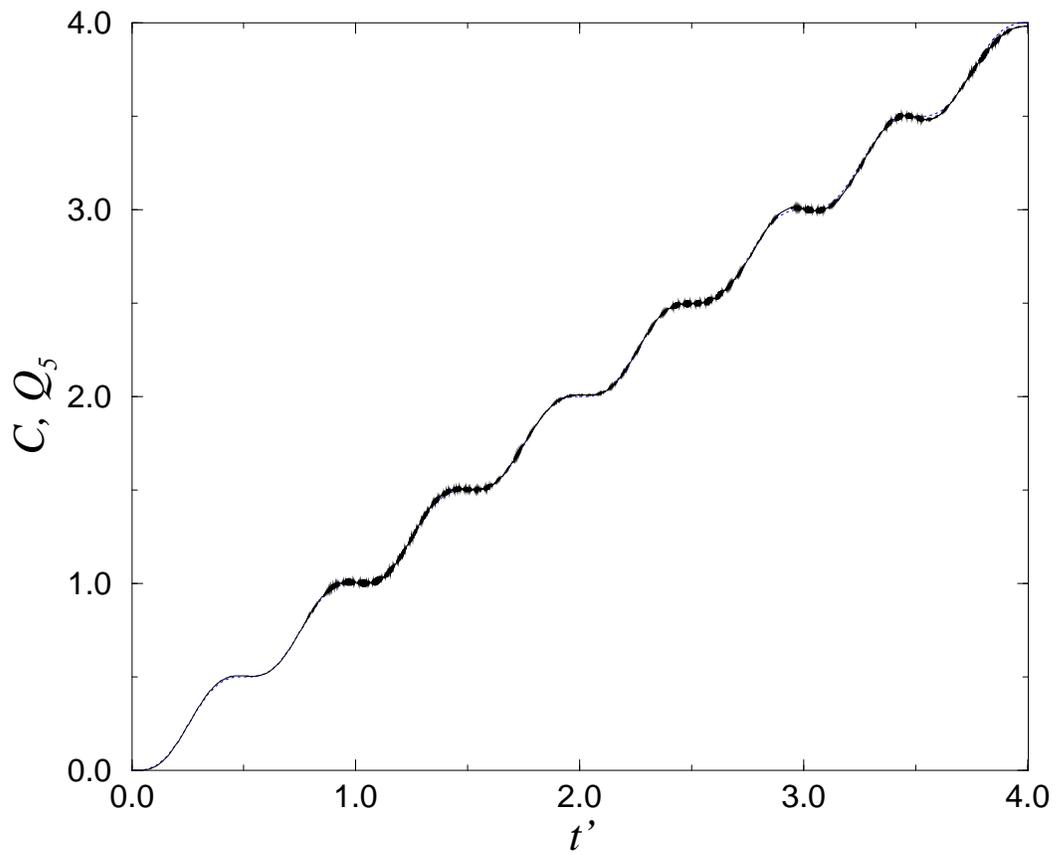,width=16.0cm}}
\caption{Response of the fermions: the axial charge $Q_5$ (solid, with 
rapid oscillations) and the Chern-Simons number $C$ (dotted) versus 
$t'$, for several sphaleron transitions, $C=0\to 4$. 
The lines fall mostly on top of each other, in accordance with the anomaly 
equation. Parameters as in (\ref{eqparhandmade}).}
\label{fighandmade2}
\end{figure}

\begin{figure}
\centerline{\psfig{figure=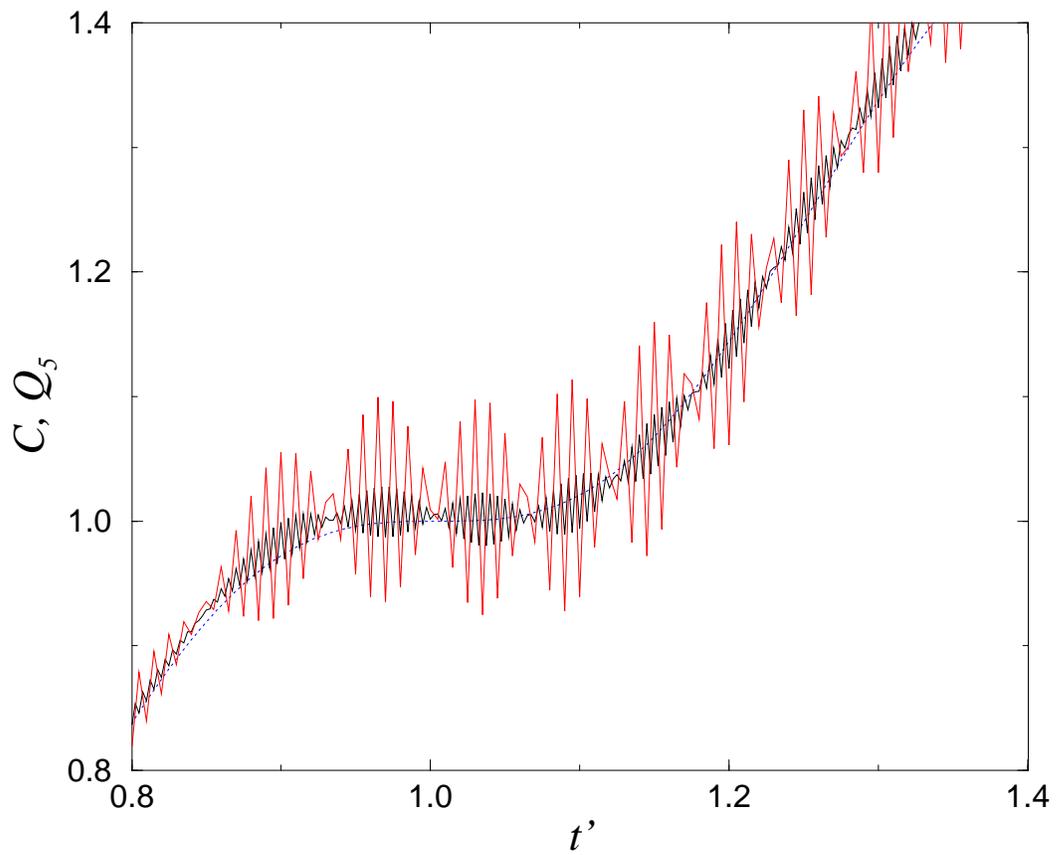,width=16.0cm}}
\caption{As in figure \ref{fighandmade2}. Blow up showing $C$ (dotted, 
slowly varying) 
and $Q_5$ with rapid oscillations $\propto a_0^2(-1)^{t/a_0}$ due to the 
time discretization, 
for two values of the lattice spacing in time, $a_0/a=0.1$ (larger 
amplitude) and $0.05$ (smaller amplitude).}  
\label{fighandmade3}
\end{figure}

 In figure \ref{fighandmade2} we show the response of the fermions, 
i.e. the change in the axial charge, when going through the series of 
configurations of figure \ref{fighandmade1}.
The axial charge follows the Chern-Simons number, which changes from 0 to 
4. This is the anomaly equation. There are high frequencies oscillations 
present in $Q_5$, which are more clear in figure \ref{fighandmade3}, 
where we 
show a blow up of figure \ref{fighandmade2} for two values of the 
lattice spacing in time, $a_0/a=0.1, 0.05$. These high frequency 
oscillations reflect that the lattice Dirac equation is a three term 
difference equation in time. Indeed, in the case of the free Dirac 
equation we find the solutions 
\[
\Psi(x,t) =  e^{ipx-i\om t}\Psi_{p\om},\;\;\;\;
\om = \pm E_p, \;\;\;\; \om = \frac{\pi}{a_0} \mp E_p.
\]
The first solution represents the normal physical particles, the second 
solution is the sum of a high frequency part ($\propto 1/a_0$) and a low 
energy part representing the doubler. The high frequency part gives a 
$(-1)^{t/a_0}$ behaviour with an amplitude that vanishes as $a_0^2$.

\begin{figure}
\centerline{\psfig{figure=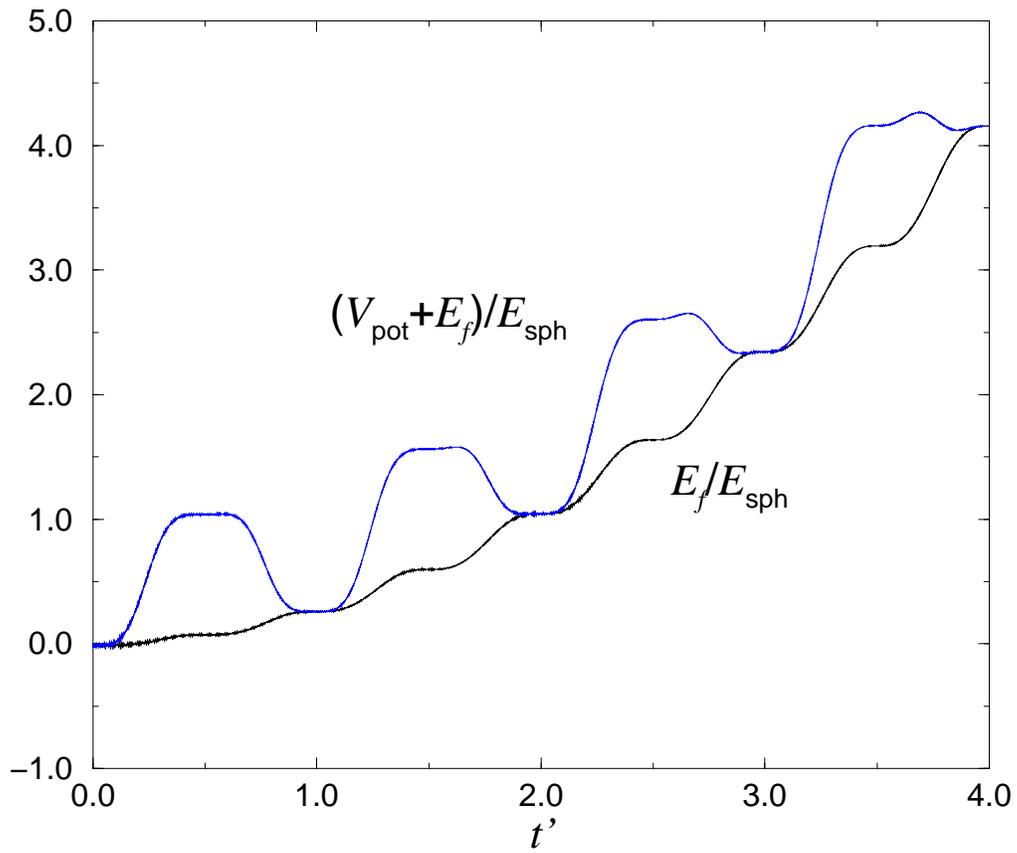,width=16.0cm}}
\caption{Energy of the fermions and the sum of the energy of the fermions and
the potential energy in the Bose fields, normalized with the 
sphaleron energy, versus $t'$. Parameters as in figure 
\ref{fighandmade2}, with $a_0/a$ = 0.005.} 
 \label{fighandmade4}
\end{figure}

Like the axial charge, the fermion energy has rapid oscillations, 
with an amplitude that decreases for decreasing $a_0/a$. 
The energy transferred to the fermions and the sum of the fermion and 
the potential energy in the Bose fields is presented in figure 
\ref{fighandmade4}. We use here a very small $a_0/a=0.005$ to suppress 
the rapid oscillations.
The fermions lift the degeneracy of the 
bosonic groundstate, and introduce local minima. The picture is symmetric 
for $C\to -C$ ($t' \to -t'$), we have no CP violation in our model.

We would like to stress that taking the doublers into account leads to a 
consistent time evolution. 
One can imagine to start with only the mode functions for the 
normal particles. Let us recall that these are initialized 
as smoothly as possible, $U_{\al u}(x,0)= U_{\al u}(x,a_0)$. Right after 
$t=0$, only the normal particles are present. However, due to the 
non-linearity of the equations of motion, high frequencies will be 
generated, and in order to make sense out of these, the doubler 
interpretation has to be introduced after all. Hence, it is better to 
include them from the start. In terms of evolution in Hilbert space, 
including the doublers explicitly leads to unitary time evolution in 
the Hilbert space that is associated with the lattice path integral.

\begin{figure}
\centerline{\psfig{figure=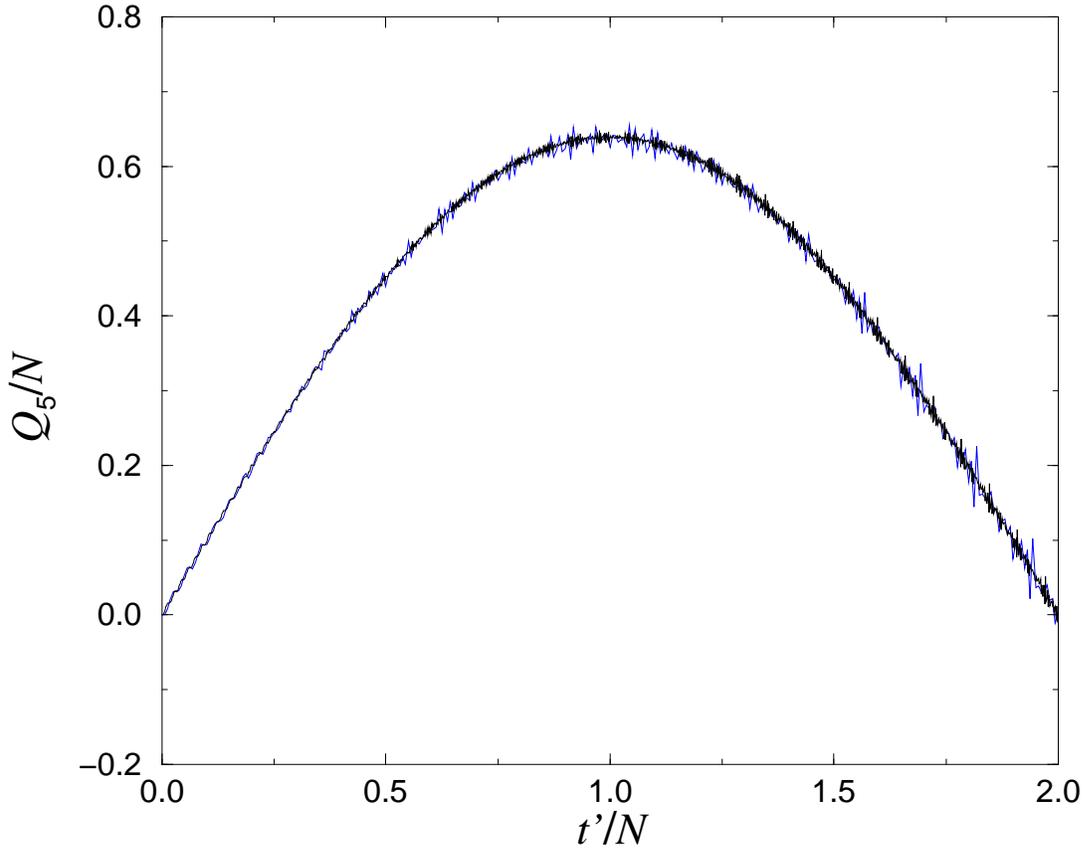,width=16.0cm}}
\caption{
Lattice periodicity of $Q_5$ for zero Yukawa coupling: 
$Q_5/N$ versus $t'/N$ for two values of the lattice spacing, $N=16, 32$. 
After the rescaling with $N$, the lines converge to the same curve, as 
expected. Other parameters as in figure \ref{fighandmade2}.} 
\label{fighandmade5}
\end{figure}

Another test is the periodicity of $Q_5$ on the lattice.
As explained in Appendix \ref{appspec} for vanishing Yukawa coupling, 
if $C=N (-N)$ the axial charge takes its maximal (minimal) 
value, and if $C=\pm 2N$, the total axial charge is zero again. 
 Although this phenomenom is a lattice artefact, which is due to the 
finite number of states, it still serves as a check on the fermion system.
We show the outcome of this test in figure \ref{fighandmade5} for $N=16, 
32$ and $G/e=0$. Plotted are $Q_5/N$ versus $t'/N$, 
because the curves should then fall on top of each other for large $N$.
We see this to be the case already for these small $N$ values.
The maximum $Q_5$ also agrees well with the value 
$2/\pi\approx 0.637$ for $N\to\infty$ (cf.\ (\ref{maxQ5})).
Of course, this maximum is way out of the region where there is
continuum behaviour,
 i.e. where the anomaly equation is satisfied ($Q_5 \approx C 
\approx t'$), 
which appears to be $|Q_5|$ smaller than $N/4$.

\begin{figure}
\centerline{\psfig{figure=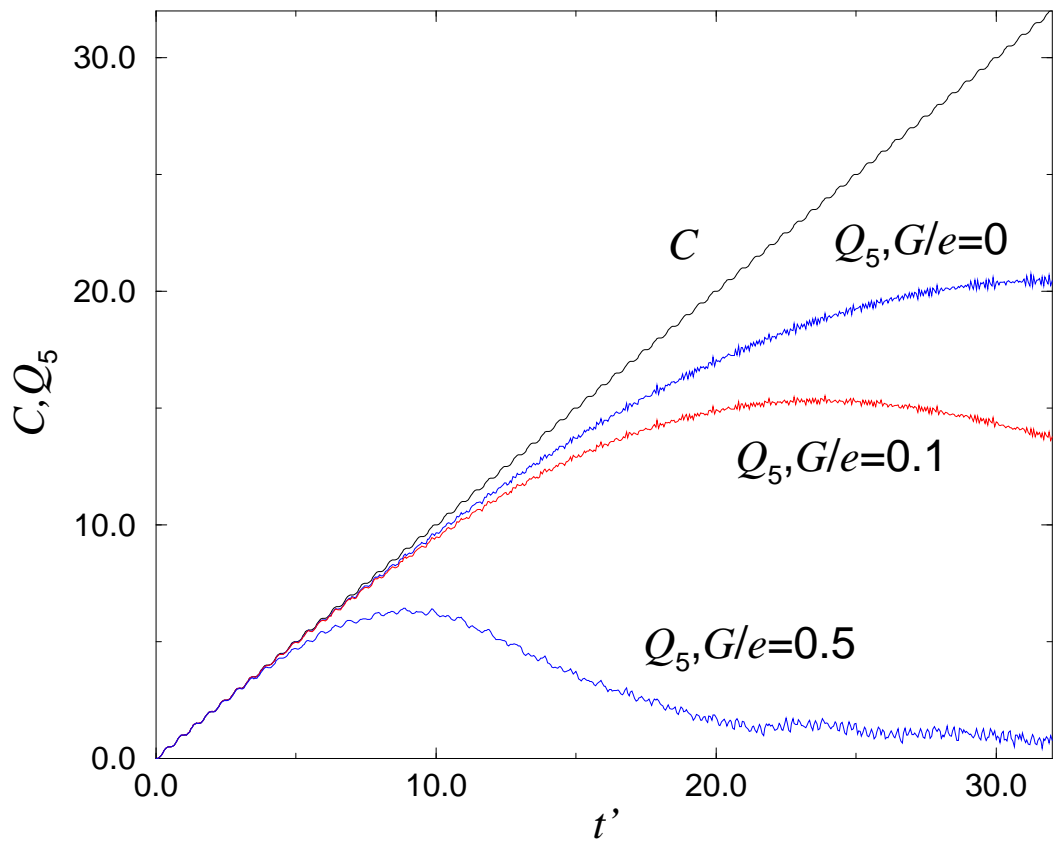,width=16.0cm}}
\caption{As in figure \ref{fighandmade2}, $C$ and $Q_5$ for three values of 
the Yukawa coupling, $G/e=0, 0.1, 0.5$. For larger Yukawa couplings, 
lattice artefacts are more important.}
\label{fighandmade6}
\end{figure}

\begin{figure}
\centerline{\psfig{figure=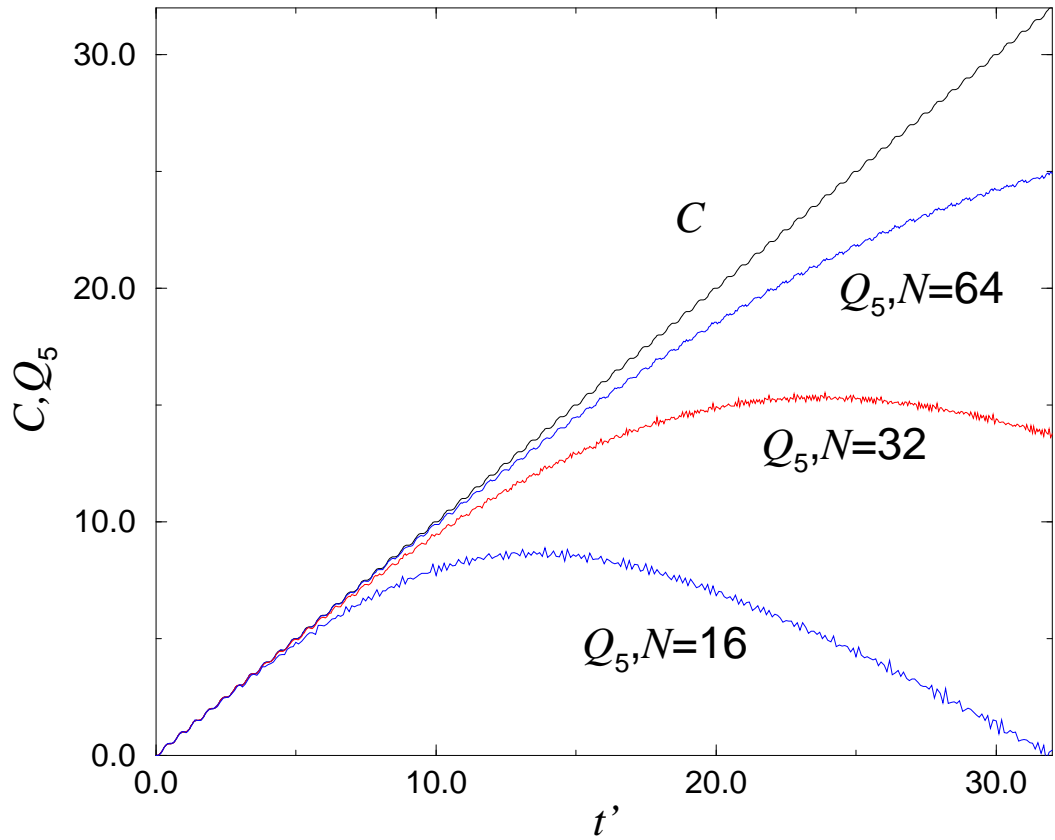,width=16.0cm}}
\caption{
As in figure \ref{fighandmade2}, $C$ and $Q_5$ for fixed Yukawa coupling, 
$G/e=0.1$, and three values of the lattice spacing, 
$N=16, 32, 64$. Decreasing the lattice spacing (increasing $N$) shows 
convergence to the continuum limit.}
\label{fighandmade7}
\end{figure}

 We proceed with an investigation of the effect of increasing the Yukawa 
coupling. 
In figure \ref{fighandmade6}, we show the behaviour of the axial charge 
for three values of the Yukawa coupling, $G/e=0, 0.1, 0.5$, up to a  large 
Chern-Simons number $C=N$.
In the continuum regime $|Q_5| \lesssim N/4=8$, the anomaly equation is 
(approximately) satisfied for finite $G$ as well, but the deviation of 
$Q_5$ from $C$ becomes more and more important for larger $G$.
This can be further investigated by taking a fixed finite 
coupling $G/e=0.1$, and decreasing the lattice spacing (increasing the 
number of lattice points). The result is shown in 
figure \ref{fighandmade7}, for $N=16, 32, 64$.
We conclude that for a finite Yukawa coupling lattice 
artefacts are more important, and that for fixed Yukawa coupling,  
an increase in the number of lattice points shows convergence to the 
continuum limit. 
It is also known from previous (lattice) studies (see e.g. \cite{Am94}) 
that a finite Yukawa coupling has a relatively large influence on the 
relation between the anomalous charge and the Chern-Simons number.

Summarizing the results obtained in this section, 
we found, using a controlled series of external Bose fields 
configurations, that the system behaves for zero and small Yukawa couplings 
as expected. For bigger Yukawa couplings, the lattice artefacts are 
substantially larger.
With this in mind, we can continue and study the complete dynamics.

\section{Renormalization}
\label{secrenorm}
\setcounter{equation}{0}

We go on with the full dynamics, and include the back reaction.
The full quantum theory contains ultraviolet divergences, which are 
regulated by the lattice cutoff. In this section, we discuss the 
renormalization and show numerical results to demonstrate its
applicability.
 
Let us first make some remarks about the role of the parameter $v^2$.
By rescaling according to 
\be
\label{rolevsq}
\ph = v\ph', \;\;\;\;
e=e'/v,\;\;\;\;
G=G'/v\;\;\;\;
\lm=\lm'/v^2,
\ee
$v^2$ appears in front of the bosonic part $S_B$ of the effective action,
with the fermionic part $-\half i\log D$  unchanged.
Hence, if we also scale the bosonic initial conditions accordingly,
the influence of the fermions is scaled by a factor $1/v^2$, and the 
fermion back reaction gets reduced at larger $v^2$.  
This can also be seen directly from the effective equations of 
motion.\footnote{To be precise, the rescaling is with a finite $v=v_R$, 
this introduces a rescaled bare $v_B'=v_B/v_R$.}
This is of course not surprising since $1/v^2$ is indeed 
the dimensionless semiclassical expansion parameter. 

As indicated in section \ref{seccontinuum}, only the scalar self energy 
is divergent in  the full quantum theory. The divergence is renormalized by 
the appropriate choice of $v^2_B$. 
Therefore, let's take a closer look at the equation for the scalar 
field (\ref{eqhiggs}). Because we treat the Bose fields as classical 
and only the fermions are integrated out, we expect to find only the 
fermion one-loop divergence of the Higgs self energy. 
To see this explicitly, we consider for simplicity a ground state 
configuration for the Bose fields, $A_1=0$, $\phi=v_R/\sqrt{2}$, where 
$v_R$ is the renormalized, effective Higgs expectation value.  
Substituting this in (\ref{eqhiggs}), gives the gap equation
\be
\label{eqgapex}
\lambda(v_R^2-v_B^2)\frac{v_R}{\sqrt{2}} = G\bra F\ket(v_R),\ee
where we indicated that the mode functions are initialized with $v=v_R$. 
Using (\ref{eqFinit}) for $\bra F \ket(v_R)$, we see that it
depends on $Gv_R$ through $m_{p\eta}$ and $E_{p\eta}$. This makes 
(\ref{eqgapex}) a highly non-linear equation in $G$ and $v_R$ .
We can, however, perform a lowest order approximation, and expand the 
right hand side in $G$, to find (at zero initial temperature $\sg_{p\eta 
f}=1$) 
\bean \bra F\ket(v_R) &=& \frac{Gv_R}{\sqrt{2}}
\frac{1}{L}\sum_{p}\frac{1}{E_{p0}}\left(1-\frac{m_{p}^2}{E_{p0}^2}\right)\\
&=& \frac{Gv_R}{\sqrt{2}}\left(\frac{1}{\pi}\log N + {\cal O}(N^0)\right),
\eean
where $E_{p0}$ is independent of $Gv_R$. The sum contains the expected 
logarithmic divergence, which we indicated as $\log N$. Using this 
approximation, the gap equation (\ref{eqgapex}) has two solutions 
\[  v_R = 0,\;\;\;\;v_R^2 = v^2_B + \frac{G^2}{\lambda}
\frac{1}{L}\sum_{p}\frac{1}{E_{p0}}\left(1-\frac{m_p^2}{E_{p0}^2}\right).\]
The log divergence in the second solution is canceled with $v_B^2$. 

In practice, we fix 
$v_R^2$ and then find (for certain $N$) the corresponding bare 
parameter $v_B^2$, simply from 
(\ref{eqgapex}),
\be
\label{eqvb}
 v_B^2 = v_R^2 - 
\frac{G}{\lambda}\frac{\sqrt{2}}{v_R}\bra F\ket(v_R),
\ee
with $\bra F\ket$ given by (\ref{eqFinit}).
We demonstrate this with numerical results. We use the following 
parameters 
\be
\label{eqparrenorm}
\begin{array}{llll}
N=16,32,48, & a_0/a=0.1, & eL=3.2, \\
v^2_R=11.15 & \lm/e^2=0.25, & G/e=0.5, 
\end{array}
\ee
and as initial Bose fields configuration, we take
\bea
\nonumber
\phi(x,0) = \frac{v_R}{\sqrt{2}},&&\;\;\;\; \partial_0\phi(x, 0) = 
3e\cos \frac{2\pi x}{L},\\
\label{eqinitrenorm}
A_1(x, 0) = 0,&&\;\;\;\; \partial_0 A_1(x,0) = \mbox{Gauss' law} + e^2.
\eea
We start with a slowly varying wave for $\partial_0\phi$. The 
space dependence 
of $\partial_0 A_1$ is determined by Gauss' law (\ref{eqgausslaw}) (it is 
zero in this simple case), only the homogeneous part can be freely chosen 
(it is the `conjugate variable' for the Chern-Simons number degree of 
freedom).
Note that for this type of initial conditions, the initial energy of 
the whole system is entirely in Bose fields with long wave lengths.

\begin{figure}
\centerline{\psfig{figure=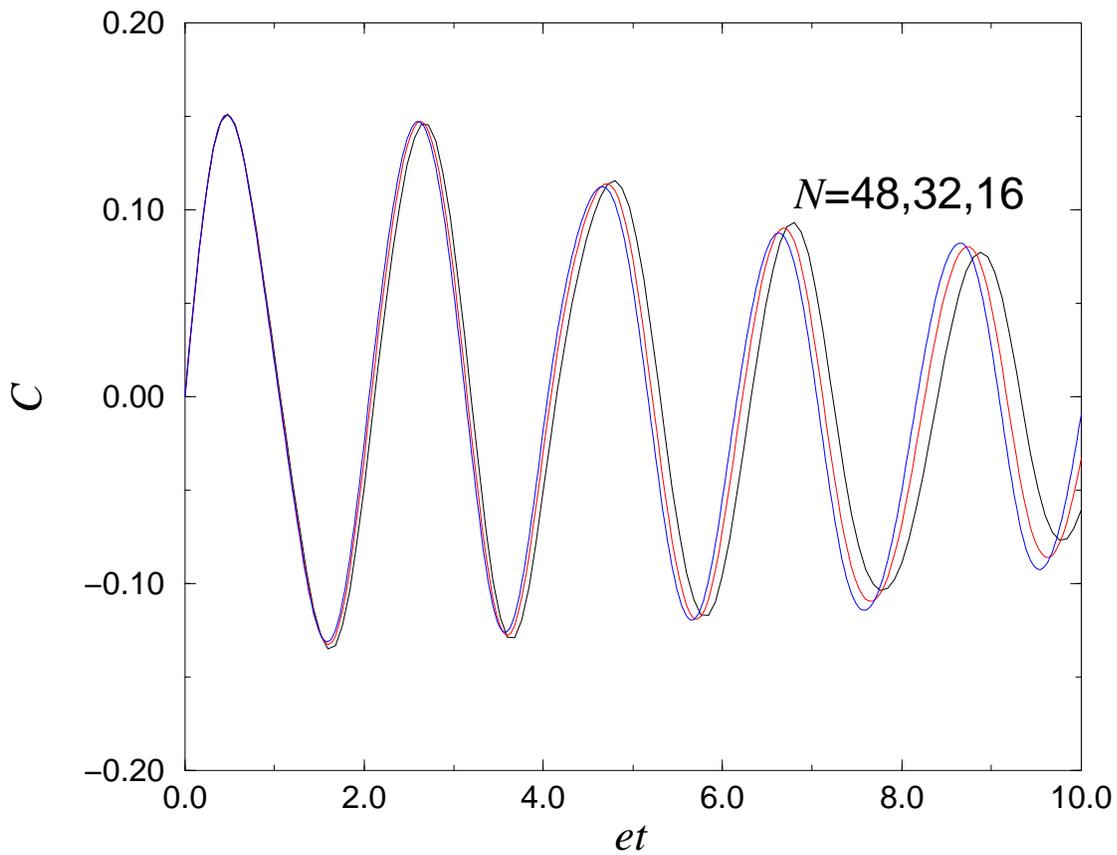,width=16.0cm}}
\caption{Need for renormalization: Chern-Simons number $C$ versus $et$ for 
fixed bare $v_B$, for three values of the lattice spacing, for 
parameters (\ref{eqparrenorm}) and initial conditions (\ref{eqinitrenorm}).
The lines show dependence on the lattice spacing.} 
\label{figrenorm1}
\end{figure}
\begin{figure}
\centerline{\psfig{figure=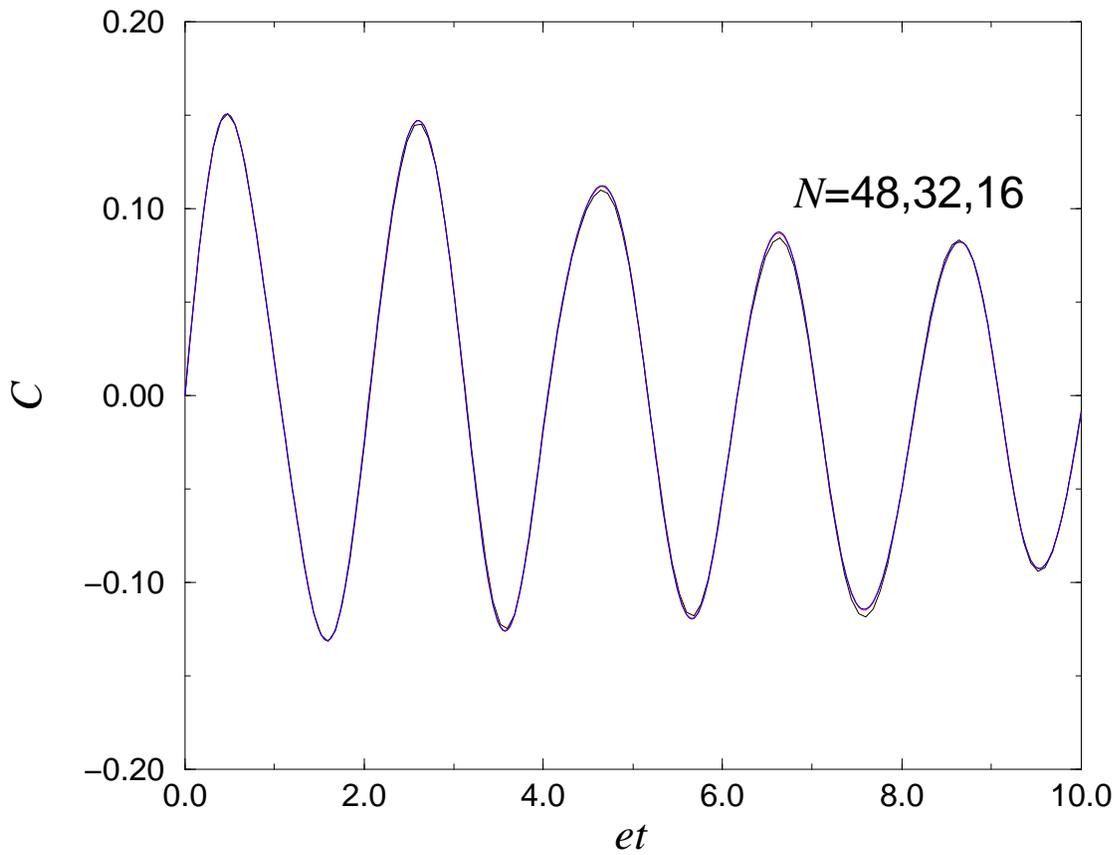,width=16.0cm}}
\caption{Renormalization: as in figure \ref{figrenorm1}, $C$ versus $et$ 
for fixed 
renormalized $v_R$, and three values of the lattice spacing. The bare 
parameter $v_B$ is determined from (\ref{eqvb}). The line with 
largest lattice spacing ($N=16$) shows the largest discrepancy, and  
convergence towards the continuum limit is clearly visible.} 
\label{figrenorm2} 
\end{figure}

In figure \ref{figrenorm1}, we show the Chern-Simons number  as a 
function of time for three values of the lattice spacing ($N=16,32,48$). 
$C$ undergoes a damped oscillation. The bare expectation value is kept 
fixed, $v_B^2=10$,  and {\em not} changed according to (\ref{eqvb}). 
The three curves do not agree, and the period is 
lattice spacing dependent.  
In figure \ref{figrenorm2}, we perform the same calculations, but now 
with a fixed renormalized expectation value, $v_R^2=11.15$. The bare 
expectation value that enters in the dynamics is determined according to 
(\ref{eqvb}), and is $N$ dependent. 
The value of $v_R$ is chosen in such a way that for $N=48$, $v_B^2=10$, 
such that the curves can be compared with the $N=48$ curve in figure 
\ref{figrenorm1}. In this case, we see that the three lines fall on 
top of each other, and there is convergence 
towards the physical  continuum limit. The line that shows the largest 
discrepancy is for largest lattice spacing ($N=16$).

\begin{figure}
\centerline{\psfig{figure=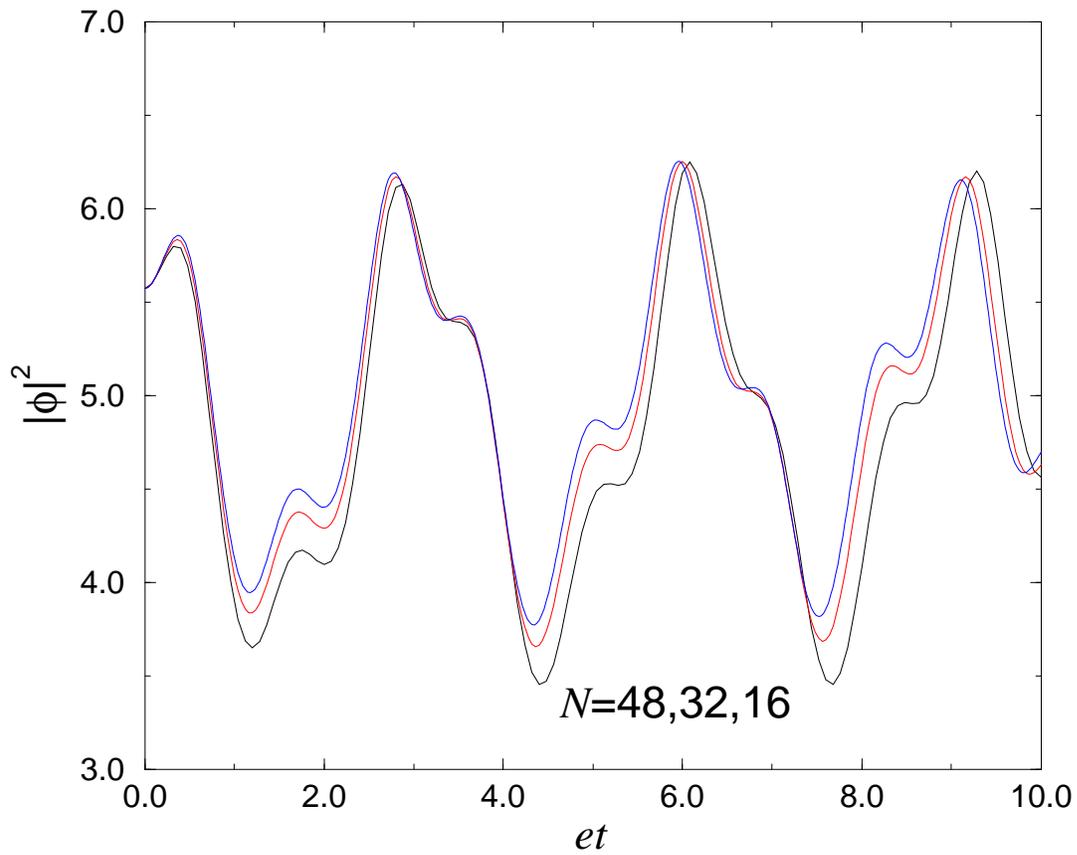,width=16cm}}
\caption{
Need for renormalization: as in figure \ref{figrenorm1}, the scalar field 
amplitude $|\ph|^2$ 
versus $et$, for fixed bare $v_B$, and three values of the lattice 
spacing. Again there is lattice spacing dependence.} 
\label{figrenorm3} \end{figure}

\begin{figure}
\centerline{\psfig{figure=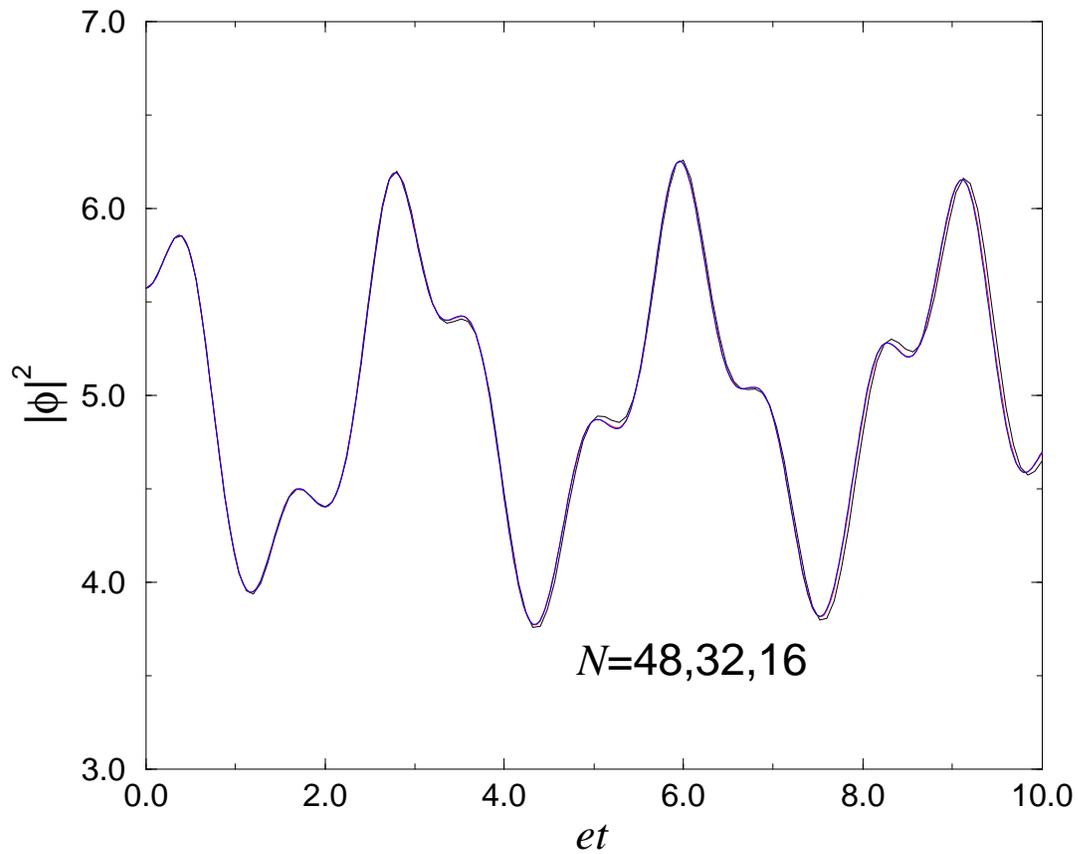,width=16cm}}
\caption{
Renormalization: as in figure \ref{figrenorm3}, $|\ph|^2$ versus $et$, for 
fixed renormalized $v_R$, and three values of the lattice spacing. 
The lines show a quick convergence towards the continuum limit, the 
lines with $N=32, 48$ are already indistinguishable.} \label{figrenorm4}
\end{figure}

To show that the observed convergence is also true for other observables, 
we repeat the analysis for $|\phi|^2= L^{-1}\sum_x |\phi(x,t)|^2$.
In figure \ref{figrenorm3}, $v_B^2$ is kept fixed, and the curves show 
again lattice spacing dependence. 
In figure \ref{figrenorm4}, we keep again $v_R^2$ fixed, and determine 
$v_B^2$ from (\ref{eqvb}). Again this leads to converging 
physical results, with the largest discrepancy for smallest $N$, $N=16$. 

We conclude that 
there is no convergence when the bare parameters are not 
adapted in the proper way, i.e. according to (\ref{eqvb}).
A correct renormalization leads to convergence towards the physical 
continuum limit, already for fairly small $N$.

\section{Non-perturbative dynamics}
\label{secnumerics}
\setcounter{equation}{0}

In the previous section, the total energy of the system was, in some sense, 
not very large. The Chern-Simons number simply 
 oscillates around zero, with an amplitude which is not larger than 0.2. 
There are no sphaleron transitions, $C$ does not get larger than 0.5.
In this section we consider systems with more energy.

To achieve this, we use the same type of initial Bose fields 
configurations as in the previous section, but with two wave lengths and 
complex amplitudes 
\bea
\nonumber
\phi(x,0) = \frac{v_{R}}{\sqrt{2}},&&\;\;\;\; \partial_0\phi(x, 0) = 
6e(1+i)\cos \frac{2\pi x}{L} +
2e(1+2i)\cos \frac{4\pi x}{L},\\
\label{eqboseinitial}
A_1(x, 0) = 0.1/L,&&\;\;\;\; \partial_0 A_1(x,0) = \mbox{Gauss' 
law} + e^2.
\eea
The parameters are
\be
\label{eqpar1}
\begin{array}{llll}
N=32, & a_0/a=0.05, & eL=3.2,\\
 v^2_{R}=8 (4), & \lm/e^2=0.25, & G/e=0.0 (0.1).
\end{array}
\ee

\begin{figure}
\centerline{\psfig{figure=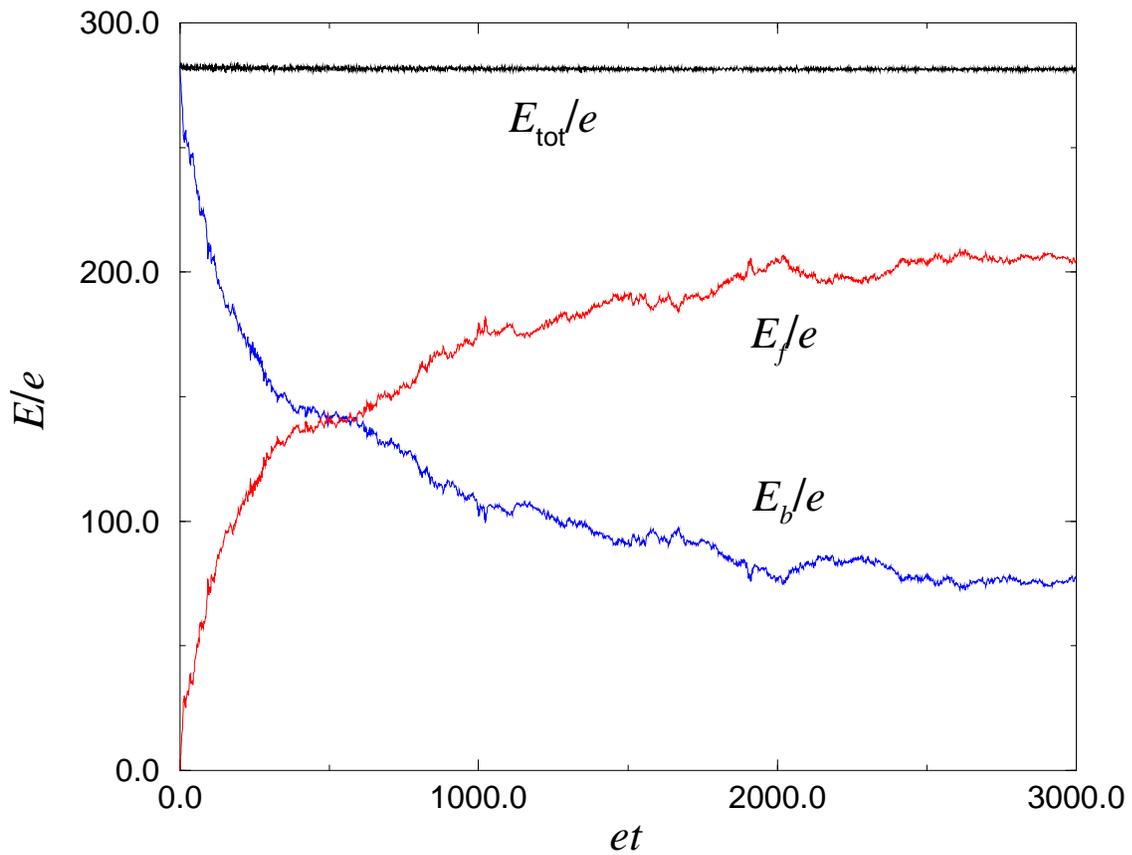,width=16.0cm}}
\caption{Non-perturbative dynamics: energy of the Bose fields $E_b/e$, the 
fermion field $E_f/e$, and the sum $E_{\rm tot}/e$ versus $et$, for 
initial conditions (\ref{eqboseinitial}) and  parameters 
(\ref{eqpar1}) with $v_R^2=8, G/e=0$.} 
\label{fignonpert1} 
\end{figure}

We begin with the case of zero Yukawa coupling. The fermions 
start in a 
vacuum state, with zero energy. All the initial energy is in the Bose 
fields. We show the energy in the Bose fields $E_b$, in the fermion field 
$E_f$ and 
the total energy $E_{\rm tot}$ 
in figure \ref{fignonpert1}, for fairly larger times 
than in the previous section, $0<et<3000$. The total energy is 
conserved up to small oscillations, very similar to the result from a 
leapfrog algorithm. The initial state is clearly a nonequilibrium 
situation, and there is energy transfer from the Bose fields to the 
fermion degrees of freedom. 
Presumably there is fermion particle production, as in the nonequilibrium 
homogeneous models  studied before \cite{pairpro}-\cite{nonequi}. 

\begin{figure}
\centerline{\psfig{figure=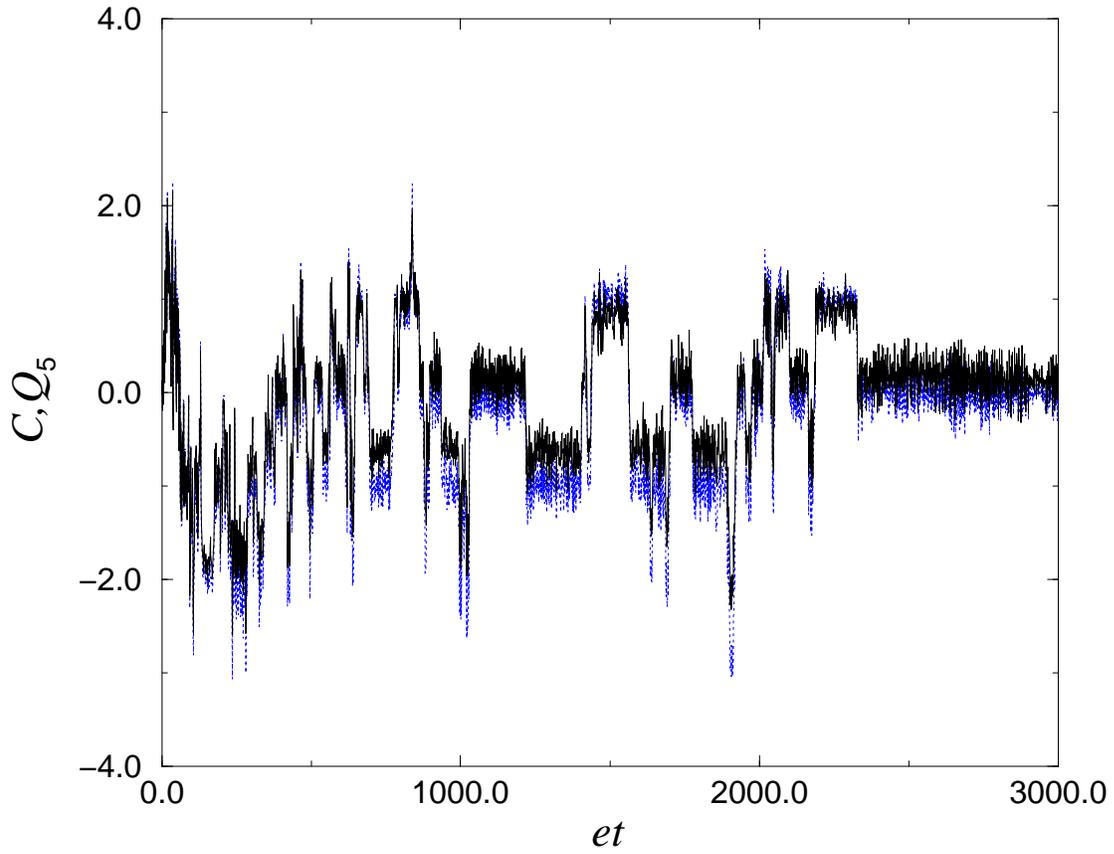,width=16.0cm}}
\caption{As in figure \ref{fignonpert1}, Chern-Simons number $C$ (dotted) 
and axial charge $Q_5$ (solid) versus $et$. The lines cannot always be 
distinguished, which is in accordance with the anomaly equation. Small 
deviations, e.g., around $et\approx 1300$, are due to the finite lattice 
spacing.} 
\label{fignonpert2} 
\end{figure}

\begin{figure}
\centerline{\psfig{figure=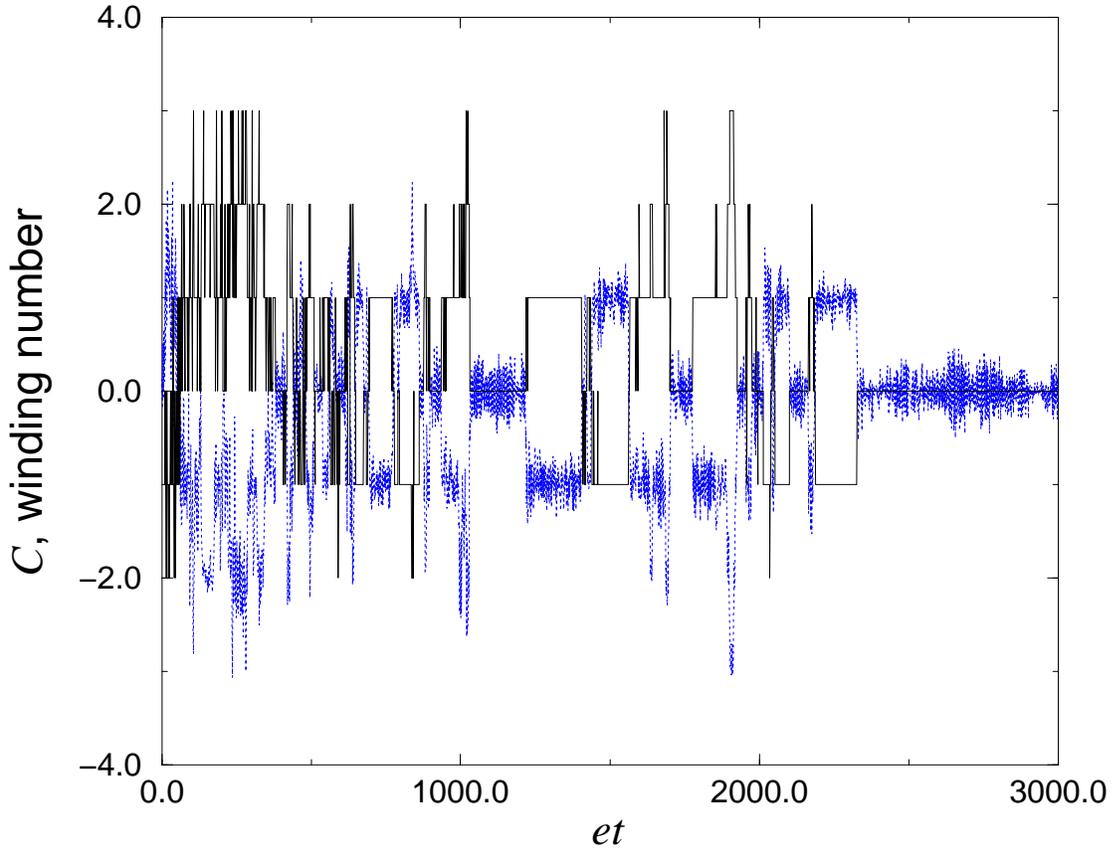,width=16.0cm}}
\caption{As in figure \ref{fignonpert1}, Chern-Simons number $C$ (dotted) 
and Higgs winding number (solid) versus $et$. 
The lines are anticorrelated, which is expected when sphaleron 
transitions dominate.}
\label{fignonpert3}
\end{figure}

In figure \ref{fignonpert2} we show the Chern-Simons 
number and the axial charge. 
Initially, there is a lot of energy present in the Bose fields. The 
Chern-Simons number is wildly moving. The Bose 
system does not feel the sphaleron energy barriers. However, there is 
continuous energy transfer towards the fermions, and plateaus become visible. 
The energy transfer continues and the time the system spends in one `vacuum' 
becomes longer. 
$Q_5$ follows $C$, 
 and it is not always possible to distinguish $Q_5$ from $C$. This is 
actually a nice property, because it shows that the anomaly equation is 
correctly reproduced.
Small differences are due to the finite lattice spacing. 
In figure \ref{fignonpert3} we show the Chern-Simons number and the Higgs 
winding number. The winding number is anticorrelated with $C$ 
when sphaleron transitions dominate, as expected from
section \ref{sechandmade}.

Let us this stage discuss the stability of the numerical algorithm. The 
total charge $Q=Q_f+Q_h$ is conserved better than ${\cal O}(10^{-10})$ 
(actually, for zero Yukawa coupling, both $Q_f$ and $Q_h$ are separately 
conserved). The first component of the vector triplet charge $Q_{\rm 
fl}$ is 
conserved better than ${\cal O}(10^{-16})$. Locally, Gauss' law is 
respected better than ${\cal O}(10^{-10})$. 

\begin{figure}
\centerline{\psfig{figure=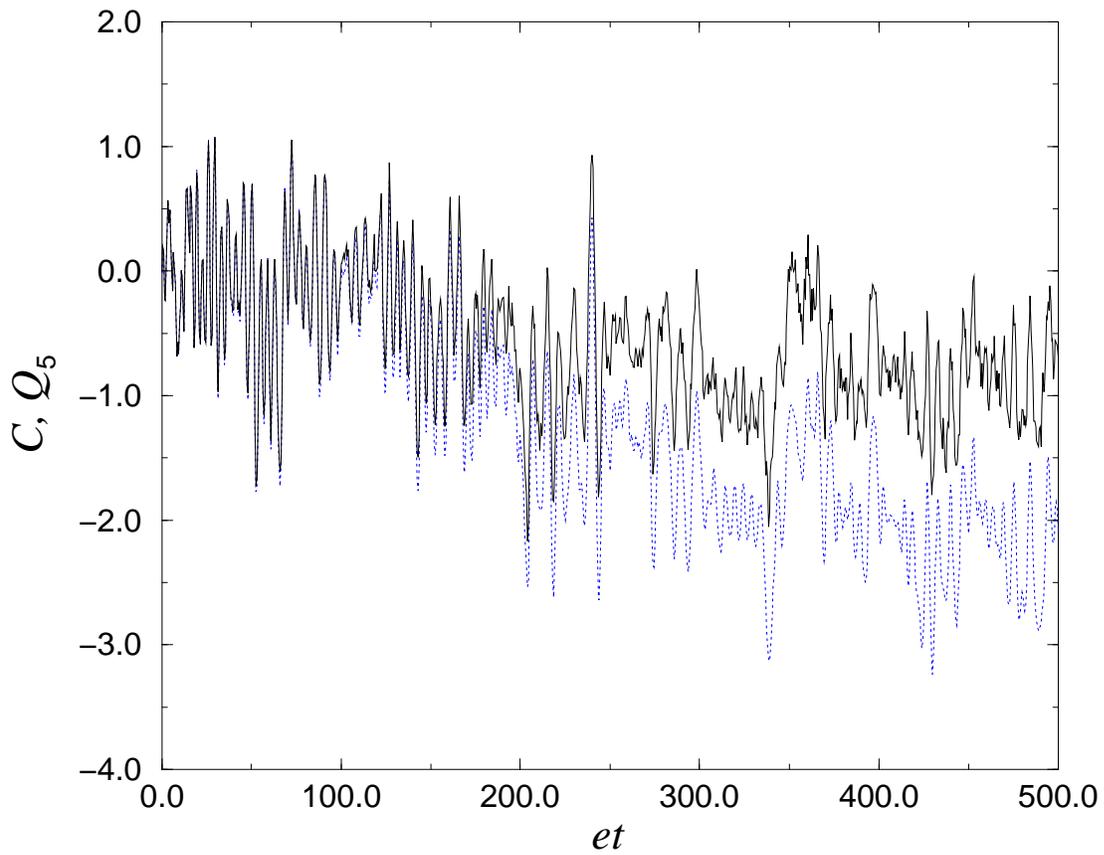,width=16.0cm}}
\caption{As in figure \ref{fignonpert1}, Chern-Simons number $C$ (dotted) 
and axial charge $Q_5$ (solid) versus $et$, for finite Yukawa coupling 
$G/e=0.1$, and $v_R^2=4$.  After a finite time, $Q_5$ looses $C$, and the 
anomaly equation is no longer satisfied.}
\label{fignonpert5}
\end{figure}

For finite Yukawa coupling, we found in section \ref{sechandmade} with 
the help of handmade sphaleron transitions, that the axial charge is more 
sensitive to the lattice discretization. 
In particular, for the typical parameters used in this section 
and with $N=32$, 
the axial charge suffers quite a lot from discretization effects.
This indeed shows up in the dynamics. In figure \ref{fignonpert5} we show 
$C$ and $Q_5$ for $v^2_R=4, G/e=0.1$ and $et$ between 0 and 500.
The anomaly equation is 
reasonably satisfied only for a finite time, in contrast to the 
$G/e=0$ example.

\section{Finite temperature and density}
\label{secthermalbose}
\setcounter{equation}{0}

Let us summarize what we have done so far. In all the results we have 
shown, the initial Bose fields are long wave length fields. The fermion 
contribution is 
calculated starting from a vacuum state, i.e. (\ref{FermiDirac}) with 
$T_{f,\rm in}=0$. It is straightforward to start with a non-zero
fermion temperature, and study e.g. the effect of (Landau) damping on the 
Bose fields due to a thermal bath of fermions. 

However, in this section we would like to comment briefly on the 
possibility of applying the methods used in this paper to a situation 
where also the Bose fields are thermal. Such a system would describe a 
thermal plasma consisting of classical Bose fields and quantum fermion 
fields. This would be interesting e.g. from the perspective of baryogenesis.

In this situation, there are several cases to be distinguished: 
either the initial state of the system may be prepared according to
a typical equilibrium configuration, or the initial state of the system 
is prepared out of equilibrium, but it reaches some sort of equilibrium 
at large times.

The first situation implies that it is known what a typical 
equilibrium situation looks like. One might think that it may be 
approximately realized as follows. 
Choose a temperature $T$ in the Fermi-Dirac distribution
(\ref{FermiDirac}) specifying the fermion initial conditions.
Prepare the Bose fields in a classical equilibrium configuration at 
this temperature, i.e.\ draw them from the classical canonical
distribution $\exp(-E_b/T)$.
Correlation functions are then computed by averaging over such 
initial configurations. Note that in this way one has prepared two 
subsystems which are {\em separately} in equilibrium, i.e. the Bose and the 
fermion subsystem, and that these are coupled together at $t>0$. 
If the theory is weakly coupled, and the back reaction of the fermions 
is perturbative, this can be analysed using perturbation 
theory, along the lines as in 3+1 \cite{BoMcSm, AaSm97} and 1+1 
\cite{TaSm98} dimensions for a purely classical Bose system.
In the classical theory (without the fermions), perturbation theory 
leads to the conclusion 
\cite{TaSm98} that the classical approximation is good\footnote{In the 
sense of the absence of ultraviolet divergences in classical thermal loops 
and the agreement with the results from high temperature quantum field 
theory.} in 1+1 dimensions,
for $T = v^3 \sqrt{\lm} T'$, with $v^2 \gg 1$ (cf.\ (\ref{rolevsq}))
and $T' = O(1)$.  
If the fermions are included, we should be able to reproduce 
numerically this way the fermion and boson propagators calculated 
analytically in the one loop approximation in the appropriate momentum 
and frequency regime.

However, an actual numerical implementation of this idea leads us to the 
conclusion 
that for times beyond the validity of the perturbative calculation, the 
fermion back reaction becomes non-perturbative and the {\em coupled} 
system is in fact still far from equilibrium. 

The other possibility, the system is initialized in a nonequilibrium 
state, but reaches some sort of equilibrium at large times,
relies on the assumption that the effective equations
of motion are chaotic in the relevant coarse grained variables, such that
at large times they lead to a distribution (by averaging over time) that
approximates the microcanonical ensemble of the original quantum system.
At present it is not known to us if this assumption is correct in some sense.
We hasten to say that the objection raised sometimes to the effective equations
that they do not incorporate scattering and therefore do not
lead to the desired equilibrium distribution, is misleading 
in the inhomogeneous case considered here.
One can easily convince oneself that there is nontrivial scattering in
the effective equations, e.g.\ by taking an initial state of two
fermion wave packets approaching each other: the localized fermions 
will simply experience approximately classical Coulomb scattering.

Still, it is a deep fundamental question what happens at large times, in
the original system, and in the effective field equations, 
given for example initially a bosonic configuration with a certain 
energy and a fermions initialized in a vacuum or thermal state. 

\begin{figure}
\centerline{\psfig{figure=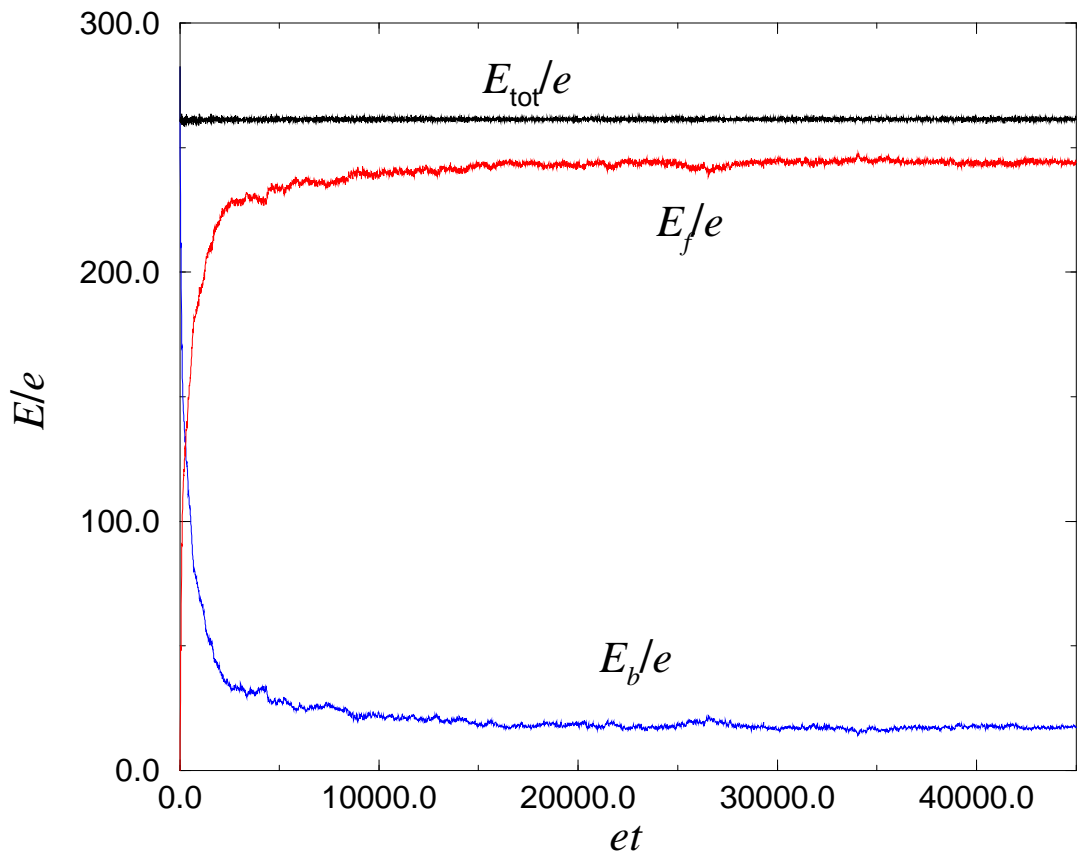,width=16.0cm}}
\caption{
Long time behaviour, energies versus $et$.
Initial conditions and parameters as in figure \ref{fignonpert1}, 
except that $G/e=0.1$.} 
\label{figlargetime}
\end{figure}

\begin{figure}
\centerline{\psfig{figure=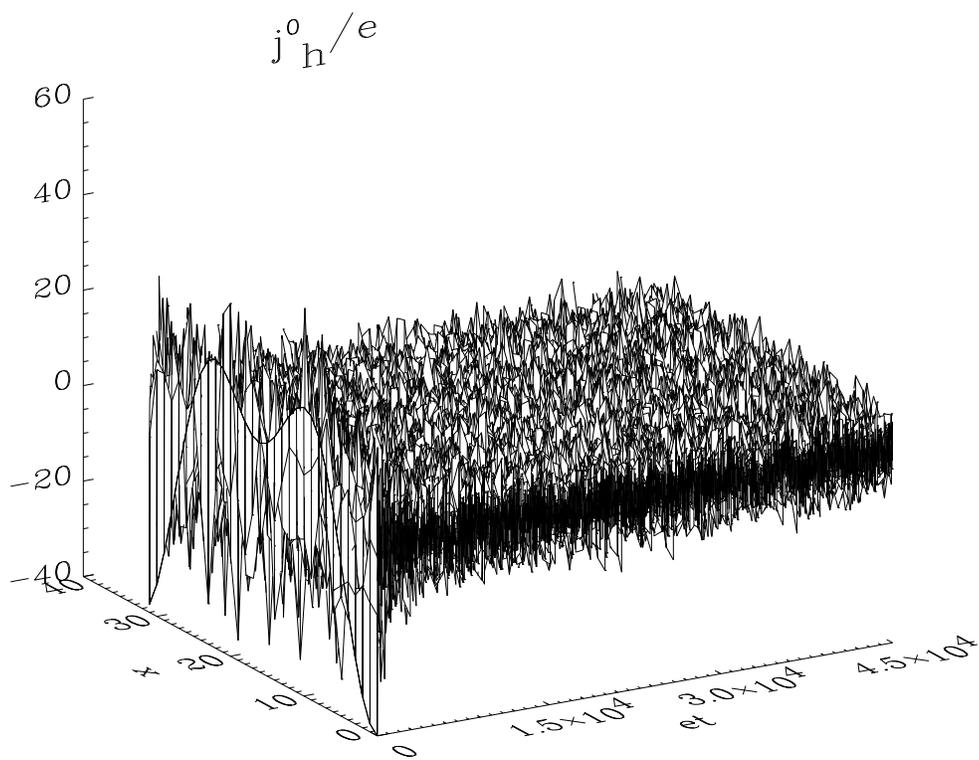,width=16.0cm}}
\caption{
As in figure \ref{figlargetime}, charge density of the scalar field 
$j^0_h(x,t)/e$ versus $x/a$ and $et$.
}
\label{figj0higgs}
\end{figure}

Here we only provide one numerical example to show the very long time 
behaviour. 
In figure \ref{figlargetime} we show the fermionic and 
bosonic energies, for initial Bose fields as in the previous section, i.e. 
(\ref{eqboseinitial}), and also a similar choice of parameters 
\[
\begin{array}{llll}
N=32, & a_0/a=0.1, & eL=3.2,\\
 v^2_{R}=8, & \lm/e^2=0.25, & G/e=0.1.
\end{array}
\] 
Notice, however, that the time interval $0<et<45000$ here is 15 times as 
large as in the previous section ($0<et<3000$).
The system is evidently initially quite out of equilibrium, while  at 
larger times the bosonic energy only drops very slowly. 
The bosonic energy drops by a factor of more than 15 as it gets
transferred to the fermions. This may be related to our neglect of the
bosonic fluctuations. 
To visualize the effect of decreasing bosonic energy in a typical bosonic 
observable, we  show in figure \ref{figj0higgs} the charge density of the 
scalar field as of function of space $x$ (in lattice units, $x/a \in 
\{0,\ldots,N-1\}$) and time $et$. Note that initially the size of the 
amplitudes decreases, but then remains approximately of the same size. 
Furthermore the system remains fluctuating, and inhomogeneous.
More detailed analysis is needed to determine if the end result is 
approximately thermal in the quantum sense.

If the system does thermalize in a reasonable way we have an
approximate nonperturbative method for computing at
finite temperature. 
An application would be  e.g. the thermal sphaleron rate in the 
presence of CP violating fermions.
Last, but not least, the same would apply to finite density. It is not 
difficult to prepare an initial state with a non-zero
conserved quantum number (e.g. $Q_{\rm fl}$ defined in section 
\ref{seclattice}). The corresponding finite density microcanonical
ensemble could then emerge at large times. It would provide a
new approach to the problem of non-perturbative finite density
computations in QCD.

\section{Summary and outlook}
\label{secconclusion}
\setcounter{equation}{0}

We considered a coupled system of classical Bose fields and a fully 
quantized fermion field, to approximate non-perturbative real-time 
dynamics in quantum field theory.
The fermion back reaction  was computed using a mode function 
expansion, similar to previous work in the homogeneous case.
However, these mode functions still contain the full space dependence and 
this makes the numerical problem scale like $N^{2d+1}$ in $d$ space 
dimensions. 
The immediate consequence is that the calculation of the back 
reaction is numerically much more demanding than in the homogeneous 
situation. 
The 1+1 dimensional model served as a useful laboratory for testing.

The lattice discretization leads to fermion 
doubling, which is well-known in euclidean lattice gauge theory.  
We added a Wilson term in space to deal with the space doublers, 
but used `naive fermions' in time. The time doublers were interpreted as a 
second flavour. In hindsight the treatment of fermion doubling could
perhaps
be done in a more elegant way, e.g.\ using staggered fermions, which
might also lead to smaller discretization errors.

We have studied several dynamical questions with the effective equations 
of motion. First we used external time dependent Bose fields to test the 
treatment of the fermions, using the mode functions. 
We found that the anomaly equation
relating the anomalous charge and Chern-Simons number is correctly 
reproduced. However, for a finite Yukawa coupling discretization errors 
are larger and qualitatively different in the sense that the anomalous
charge looses Chern-Simons number after some time. This time increases
with decreasing lattice distance.
 
Next we studied the quantum divergence which is present in the 
effective equations of motion and we showed how this can be renormalized. 
We then demonstrated the applicability of the approximation 
to nonequilibrium time evolution on a few initial value problems. 
Finally, we have discussed the possibility that the effective dynamics  
can be used to describe a plasma at finite temperature and density. 
We have shown some preliminary results at very long times which indicated
equilibration to an ensemble of (inhomogeneous) configurations.

Possible directions for the future are many.
CP violation can be incorporated by extending the model with more fermion 
fields, relevant for baryogenesis. 
Analytical calculations of various damping mechanisms of Bose fields due 
to a thermal bath of fermions can be checked numerically. 
In the typical examples we have shown, there is a large energy transfer 
from the Bose fields to the fermion fluctuations, as in  inflationary 
scenarios. This energy transfer can presumably
be interpreted as particle production, 
using the idea of adiabatic particle numbers.   
An analysis of this would require a space dependent Bogoliubov 
transformation. 
All of this is on reasonably firm theoretical grounds. Concerning the long 
time behaviour, and the possible approach to (some) equilibrium, we would 
like to stress that many questions are still open and require a detailed study.

A big issue is of course how to extend this approach to 3+1 dimensions. 
Brute force using all the mode functions seems hopeless, 
but useful inspiration can perhaps be found in the dynamical fermion 
methods used in four dimensional euclidean lattice QCD.
It should of course also be kept in mind that the ultraviolet properties of 
classical gauge fields in thermal equilibrium are different in $3+1$ 
dimensions than in $1+1$ dimensions.

\subsubsection*{Acknowledgments}  
 
It is a pleasure to thank Gerard Barkema and Bert-Jan Nauta for useful 
discussions and comments. This work is supported by FOM.

\renewcommand{\thesection}{\Alph{section}}
\setcounter{section}{0}
\renewcommand{\theequation}{\Alph{section}.\arabic{equation}}
 
\section{Dirac matrices, Dirac and Majorana spinors}
\label{appconventions}
\setcounter{equation}{0}

Our conventions are the following. The Dirac matrices obey $\{\gm^\mu, 
\gm^\nu\}=g^{\mu\nu}$, with $-g^{00}=g^{11}=1$. The antisymmetric 
Levi-Civita tensor $\ep^{\mu\nu}$ is defined by $\ep_{01}=1$. 
$\gm_5=-\gm^0\gm^1$.  
The charge conjugation matrix obeys ${\cal C}^\dagger =
{\cal C}^{-1}, {\cal C}^T=-{\cal C}$, and 
${\cal C}{\gm^\mu}^T{\cal C}^\dagger = -\gm^\mu$, and hence 
\[  {\cal C}{\gm_5}^T{\cal C}^\dagger = - \gm_5,
\;\;\;\; {\cal C}P_{L,R}^T{\cal C}^\dagger = P_{R,L},
\]
which is different than in $3+1$ dimensions. We also use the hermitian 
matrices $\al^1 = -\gm^0\gm^1, \bt = i\gm^0$. The adjoint spinor is 
$\bar\psi=\psi^\dagger\beta$.

As  explicit representation we use the Majorana-Weyl representation 
$\gm^0 = -i\sigma_2, \gm^1=\sigma_1, \gm_5 = \sigma_3$, and ${\cal C}=\bt$.

Working on the lattice with Wilson fermions, the following combination 
often appears \[  P_\pm^\mu = \half(r_\mu \pm \gm^\mu).\]
In this paper we restrict ourselves to $r_0=0$.

Finally, Pauli matrices acting on the 1,2 components of a Majorana spinor 
are denoted with $\rho_{1,2,3}$ and Pauli matrices acting in flavour 
space with $\tau_{1,2,3}$.

\section{Eigenspinors in a vacuum background}
\label{appinitialf}
\setcounter{equation}{0}

In this paper we take as initial mode functions the eigenspinors of the 
Dirac hamiltonian in a vacuum background of Bose fields, i.e. 
$\phi=v_R/\sqrt{2}, A_1=0$. In this case the 
eigenspinors can be found analytically, by spatial Fourier 
transformation. 

The Dirac hamiltonian (\ref{eqHDfl}) in a vacuum background is given in 
momentum space by \be
\HD_{{\rm fl}p} = \alpha^1 s_p + \beta (m_p + \rho_3\tau_3 m_F), \ee
with $s_p = a^{-1}\sin ap$, the spatial Wilson mass $m_p = 
a^{-1}r_1(1-\cos ap)$, and $m_F = Gv_R/\sqrt{2}$. $\rho_3$ acts on the 
Majorana components, and $\tau_3$ acts in flavour space. 
We use antiperiodic boundary conditions for the fermion field. This 
quantizes $p$ as 
\be
\label{eqp}
   p=\frac{2\pi}{L}(n-\half), \;\;\;\;n\in\{ -\half N+1, \ldots, \half 
N\},
\ee
assuming $N$ is even.
The positive energy eigenvalues are summarized by
\[
E_{p\eta} = \sqrt{ s_p^2 + m_{p\eta}^2}, \;\;\;\;
m_{p\eta} = m_p + \eta m_F,
\;\;\;\;\eta=\pm.
\]
$\eta$ refers to the product of the eigenvalues of $\rho_3$ and 
$\tau_3$. The mixing of the Wilson mass with the 
mass due to the Higgs vacuum expectation value gives a mass splitting 
and lifts the degeneracy.

The eigenspinors can be written conveniently using the 2-spinor
\[ u_{p\eta} = 
\frac{1}{\sqrt{2E_{p\eta}(E_{p\eta}-s_p)}}\left( \begin{array}{c}
 m_{p\eta} \\  i(E_{p\eta}-s_p) \\
\end{array}\right),\]
which is normalized to one, $u^\dagger_{p\eta}u_{p\eta}=1$.
 
We denote the positive energy Majorana eigenspinors, with energy 
$E_{p\eta}$ and flavours $u,d$,
with $U_{p\eta u}$ and $U_{p\eta d}$. They are easily constructed,
\[
U_{p+u} = \left(\begin{array}{c} 
u_{p+}\\
0
\end{array}
\right),
\;\;\;\;
U_{p-u} = \left(\begin{array}{c} 
0\\
u_{p-}
\end{array}
\right),
\;\;\;\;
U_{p+d} = \left(\begin{array}{c} 
0\\
u_{p+}
\end{array}
\right),
\;\;\;\;
U_{p-d} = \left(\begin{array}{c} 
u_{p-}\\
0
\end{array}
\right).
\]
The eigenspinors needed in the mode function expansion in section 
\ref{seclattice} are then given by
\be
\tilde U_{\al u}(x)  = \frac{e^{ipx}}{\sqrt{L}}U_{p\eta u}, \;\;\;\;
\tilde U_{\al d}(x)  = \frac{e^{ipx}}{\sqrt{L}}U_{p\eta d}.
\ee
where we used the collective label $\al = (p, \eta)$.
The initial conditions for the  mode functions in section \ref{secequations} 
are given by (\ref{eqinitialU}).

\section{Spectral flow}
\label{appspec}
\setcounter{equation}{0}

The idea underlying spectral flow is that the equations of motion for 
the fermions are solved in the presence of external fields, 
in the special case that these external fields are 
extremely slowly varying. Then it is possible to use the adiabatic theorem 
\cite{Mess} and determine the time evolution by diagonalization of the 
Dirac hamiltonian for each configuration of external fields (instead of 
solving the equations explicitly).

In this appendix we discuss the spectral flow analysis, for the special 
case of zero Yukawa coupling and a homogeneous gauge field.
We use Dirac fermions, with the action (\ref{eqactionqed}) and
$G=0$. In this case 
the spectrum can be found analytically for each value of the gauge field.
The difference with the real-time analysis in the main part of this 
paper, is that the spectrum is calculated instantaneously for a given gauge 
field configuration. Hence the time doublers do not appear. 
The doublers in space are lifted due to the Wilson term. We use lattice 
units $a=1$ in this appendix.

Let's start with the free ($A_1=0$) Dirac hamiltonian.
In momentum space, it is given by
\[  \HD_p = \al^1s_p +\beta m_p,\]
using the notation from appendix \ref{appinitialf}.
The eigenvalues and eigenspinors are 
\[  \HD_pw_{p\ep} = \ep E_pw_{p\ep}, \;\;\;\;
E_p = \sqrt{ s_p^2 + m_p^2}, \;\;\;\;\ep=\pm,\]
with
\[ w_{p\ep} = 
\frac{1}{\sqrt{2E_{p}(E_{p}-\ep s_p)}}\left( \begin{array}{c}
 m_{p} \\  i(\ep E_{p}-s_p) \\
\end{array}\right),\;\;\;\;
w^\dagger_{p\ep}w_{p\ep'} = \delta_{\ep\ep'}.\] 
Due to the Wilson mass term, these eigenspinors are not 
eigenspinors of $\gm_5$ with eigenvalues $\pm 1$. Instead they obey
\be 
\label{eqspec1}
w^\dagger_{p\ep}\gm_5w_{p\ep'}=\ep\frac{s_p}{E_p}\delta_{\ep\ep'},
\ee
where $s_p/E_p$ lies between -1 and +1. The sign, however, can still be 
used to indicate the would-be chirality in the continuum.

\begin{figure}
\centerline{\psfig{figure=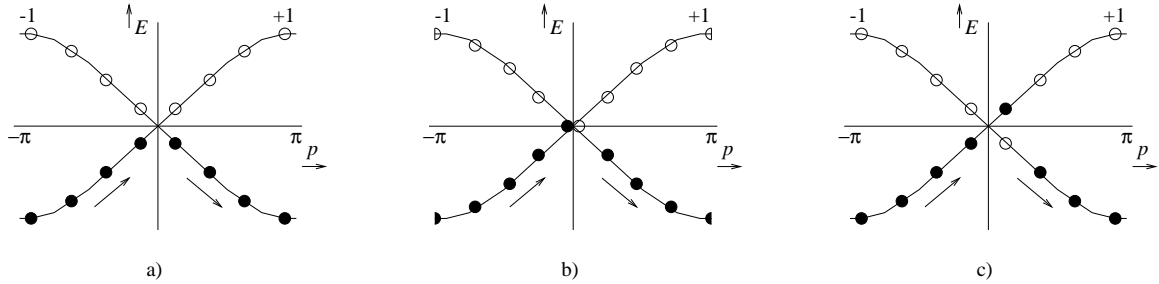,height=3.7cm}}
\caption{Spectrum of the Dirac hamiltonian and distribution of 
occupied states on a spatial lattice with $N=8$ sites, for a)~$C=0$, the 
filled Dirac sea, b)~$C=1$, 
c)~$C=2$. The occupied states are indicated with a full dot, the empty 
states with a open dot. The response to an increasing Chern-Simons 
number is indicated with an arrow. The $\pm 1$ denotes the sign of the 
$\gm_5$ expectation value of the states on the particular branch.} 
\label{figspec1}
\end{figure}

The fermion field is expanded as\footnote{Usually the $w$'s and $a$'s are 
renamed as 
$w_{p+} = u_p, w_{-p-} =v_p, a_{p+} = b_p, a_{-p-} = d^\dagger_p$.}
\[ \psi(x) = \frac{1}{\sqrt N}\sum_{p\ep} e^{ipx}a_{p\ep}w_{p\ep},\]
where the creation and annihilation operators obey the usual 
anticommutation relations
\[  \{ a_{p\ep}, a^\dagger_{p'\ep'} \} = \delta_{pp'}\delta_{\ep\ep'},\]
and zero for the other ones.
The vacuum, i.e. the Dirac sea, is given by the non-zero expectation 
values
\[  
\bra a_{p+} a_{p+}^\dagger\ket = \bra a_{p-}^\dagger a_{p-}\ket = 1.
\]
In figure \ref{figspec1}a we show the spectrum and the filled Dirac sea for 
$N=8$.
The total (bare) energy and the axial charge are given by
\bea
\label{eqappeq5}
  E &=& \sum_x \bra\psi^\dagger \HD\psi\ket = -\sum_p E_p,\\
Q_5 &=& \sum_x \bra\psi^\dagger\gm_5\psi\ket = -\sum_p\frac{s_p}{E_p} = 0.
\eea
The total axial charge is zero for the Dirac sea because of the precise 
cancellation between the $+$ and the $-$ branch. Because of the 
antiperiodic boundary conditions, there is no exact zero energy state.

The case of 
a homogeneous gauge field is now simply found by 
minimal substitution
\[  p=\frac{2\pi}{N}(n-\half)\;\;\; \to\;\;\;   p-\half A_1 
=\frac{2\pi}{N}(n-\half + \half C), \]
where we use that the Chern-Simons number $C = -NA_1/2\pi$. Recall that 
the fermions have charge $q=\half$.
In figure \ref{figspec1}, we show the spectrum for three values of the 
Chern-Simons number. For $C=1$, the spectrum is not identical to the 
spectrum at $C=0$. There are now two states with exactly zero energy (one 
filled and one empty), and there are two states exactly at the boundary 
of the Brillouin zone ($p=\pm\pi$). 
This is not contradictory, $C+1$ is not gauge equivalent to $C$, due to 
the $q=\half$ charged fermions. $C+2$ is gauge equivalent to $C$, and 
indeed, the spectrum shown in figure \ref{figspec1}c for $C=2$ is 
identical to the original one with $C=0$.
Only the occupied and unoccupied states have been shuffled.

The precise way in which this shuffling is done cannot be found using 
only the instantaneous spectrum. However, in the adiabatic 
approximation the Chern-Simons number has an implicit time dependence,
and this gives a relation between the (un)occupied states for different 
values of $C$, by continuity.
For instance, states that seem to disappear from the spectrum at $(p=\pm 
\pi, E)$ return at $(p =\mp \pi, E)$, because of the periodicity of the 
lattice. 
A calculation of $Q_5$ shows that the 
contributions from the $+$ and the $-$ branch no longer cancel, and we find
\[  Q_5(C) - Q_5(0) \approx C - 0,\]
which is of course the anomaly equation.
Here it is important that states that are close to the continuum, i.e. with 
$|E_p|\ll 1$, do have an axial charge of approximately $\pm 1$. 
This is not the case for states deep in the Dirac sea (see Eq. 
(\ref{eqspec1})).

\begin{figure} 
\centerline{\psfig{figure=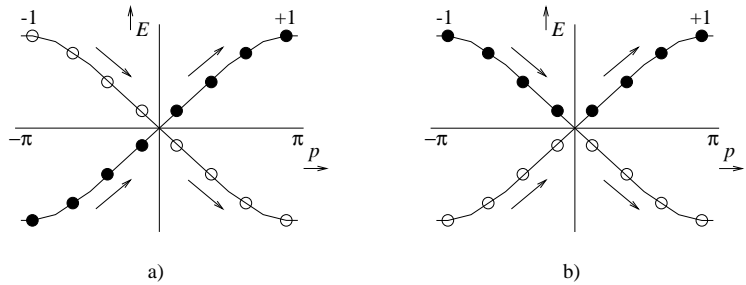,height=3.7cm}}
\caption{As in figure \ref{figspec1}, for a)~$C=N$, with
maximal $Q_5$, b)~$C=2N$, the filled Dirac sky.} 
\label{figspec2}
\end{figure}

We continue and look at larger Chern-Simons numbers. Since we only have a 
finite number of states, at 
some point all the occupied states have moved to the $+$ branch. This is 
shown in figure \ref{figspec2}a. In our model this happens when $C=N$. 
This particular distribution of occupied states has the largest axial 
charge that is possible in the lattice model, and it is given by
\be
\label{maxQ5}
\frac{Q_{5\rm max}}{N} = \frac{1}{N}\sum_p\frac{|s_p|}{E_p} 
\stackrel{N\to\infty}{\longrightarrow} 2\int_0^\pi \frac{dp}{2\pi} 
\frac{s_p}{E_p} = \frac{2}{\pi}, 
\ee
where the last expression is valid in the
limit $N\to \infty$.
When we continue to even larger Chern-Simons numbers, all the 
occupied states end up in the 
Dirac sky, and the Dirac sea has completely dried up, as shown in figure 
\ref{figspec2}b. The axial charge is zero again.
 
It is clear that for these large values of the Chern-Simons number, 
lattice artefacts are substantial.

\newpage

\hspace{0cm}

\hspace{0cm}

\hspace{0cm}

\hspace{0cm}

\hspace{0cm}

\hspace{0cm}

\hspace{0cm}

\hspace{0cm}

\hspace{0cm}

\hspace{0cm}

\hspace{0cm}

\hspace{0cm}

\hspace{0cm}

\hspace{0cm}

\hspace{0cm}

\newpage

\vspace{3cm}

\end{document}